\documentclass[
reprint,
aps,
prb,
amsmath,amssymb,
floatfix,
]{revtex4-2}

\usepackage{graphicx}
\usepackage{dcolumn}
\usepackage{soul}
\usepackage{siunitx}
\usepackage{caption}
\usepackage{color}
\usepackage{bm}
\usepackage{cancel}
\usepackage{booktabs}
\usepackage{braket}
\usepackage{mathrsfs}

\usepackage{tabularx,booktabs,caption}
\newcolumntype{Z}{>{\raggedright}X}

\DeclareSymbolFont{cmbrightop}{OT1}{cmbr}{m}{n}
\DeclareMathSymbol{\sfPsi}{\mathalpha}{cmbrightop}{9}
\begin{document}


\title{Symmetries in zero and finite center-of-mass momenta excitons}

\author{Robin Bajaj$^1$, Namana Venkatareddy$^1$, H. R. Krishnamurthy$^{1,2}$}
\author{Manish Jain$^1$}
\email{mjain@iisc.ac.in}

	\affiliation{$^1$Centre for Condensed Matter Theory, Department of Physics, Indian Institute of Science, Bangalore 560012, India }
    \affiliation{$^2$International Centre for Theoretical Sciences, Tata Institute of Fundamental Research, Bengaluru 560089, India}

\date{\today}

\begin{abstract}
We present a symmetry-based framework for the analysis of excitonic states, incorporating both time-reversal and space-group symmetries. We demonstrate the use of time-reversal and space-group symmetries to obtain exciton eigenstates at symmetry-related center-of-mass momenta in the entire Brillouin zone from eigenstates calculated for center-of-mass momenta in the irreducible Brillouin zone. Furthermore, by explicitly calculating the irreducible representations of the little groups, we classify excitons according to their symmetry properties across the Brillouin zone. Using projection operators, we construct symmetry-adapted linear combinations of electron-hole product states, which block diagonalize the Bethe–Salpeter equation (BSE) Hamiltonian at both zero and finite exciton center-of-mass momenta. This enables a transparent organization of excitonic states and provides direct access to their degeneracies, selection rules, and symmetry-protected features. As a demonstration, we apply this formalism to monolayer MoS$_2$, where the classification of excitonic irreducible representations and the block structure of the BSE Hamiltonian show excellent agreement with compatibility relations derived from group theory. Beyond this material-specific example, the framework offers a general and conceptually rigorous approach to the symmetry classification of excitons, enabling significant reductions in computational cost for optical spectra, exciton–phonon interactions, and excitonic band structure calculations across a wide range of materials.

\end{abstract}

\maketitle

\section{\label{sec:level1}INTRODUCTION}
Symmetry principles lie at the heart of quantum physics, governing both fundamental laws and emergent phenomena. In quantum systems, the transformation properties of energy eigenstates under symmetry operations \cite{wigner1931gruppentheorie,Dresselhaus2008Group,Tinkham2003Group,Bassani1975Electronic} dictate degeneracies, selection rules, and responses to external perturbations. In electronic structure calculations, for instance, Bloch's theorem provides a powerful simplification by exploiting the lattice periodicity, restricting calculations to momenta within the first Brillouin zone (BZ). Additionally, space group symmetries allow for further computational efficiency by identifying symmetry-equivalent points in the BZ, reducing the sampling space and enabling the classification of eigenstates into irreducible representations. These group-theoretical techniques are routinely employed in both one-electron and lattice dynamical problems to streamline band structure calculations and enforce optical selection rules.

In one-electron as well as phonon band structures, eigenstates at a given point in the BZ are routinely labeled by the irreducible representations of the little group at that point giving the symmetry classification of first-principles wave functions~\cite{Dresselhaus2008Group,Tinkham2003Group,Bassani1975Electronic}. These symmetry-based insights are now deeply integrated into modern electronic structure workflows, providing both conceptual guidance and computational gains across condensed matter theory. Tools such as the Bilbao Crystallographic Server~\cite{61e2fc5e1db4736e1e97ef6d}, \textsc{spgrep} package~\cite{Shinohara_spgrep_On-the-fly_generator_2023}, and \textsc{IrReP} package~\cite{IRAOLA2022108226} enable block diagonalization of Hamiltonians, degeneracy classification, and compatibility relation tracking capabilities that are now central to high-throughput and symmetry-aware materials discovery.
 
Despite their success in one-electron and phonon problems, such symmetry-based approaches remain underutilized in the context of excitons~\cite{DESLIPPE20121269,MARINI20091392}. Excitons feature prominently in the optical response of semiconductors and insulators~\cite{PhysRevB.90.205422,PhysRevLett.113.076802} and are particularly well described within the Bethe–Salpeter equation (BSE) formalism~\cite{RevModPhys.74.601,PhysRevB.62.4927,PhysRevLett.80.4510,PhysRevLett.81.2312}.  
The eigenstates of the BSE Hamiltonian, being two-particle bound states labeled by their center-of-mass (c.m.) momenta and dependent on relative momenta, differ from one-electron eigenstates in terms of the structure and the application of symmetry operations. Recently, some studies have addressed the role of symmetry in excitons. For example, Reference~\cite{PhysRevB.91.075310} analyzes the excitonic band structure of monolayer MoS$_2$ and interprets the symmetry of low-lying excitons using group theory. Ref.~\cite{nature_mos2} reports measurements of the exciton fine structure in monolayer MoS$_2$, highlightling the irreducible representations of two optically active (bright) excitons with parallel spins, along with two spin-forbidden dark states. Galvani \emph{et al.}~\cite{PhysRevB.94.125303} investigate the symmetry properties of excitons in a monolayer hBN by combining \emph{ab initio} calculations with a tight-binding Wannier analysis in both real and reciprocal space. Similarly, Ref.~\cite{PhysRevB.107.045430} examines the influence of uniaxial strain on the symmetry classification of excitons in C$_3$N, employing a tight-binding BSE framework. In another work focused on excitonic $g$ factors in monolayer WSe$_2$~\cite{junior2025generalizedmanybodyexcitongfactors}, the low-energy excitons are classified by tracking the compatibility relations between the little group $C_{3h}$ at the K/K' points and the full point group $D_{3h}$ at the $\Gamma$ point. Moreover, several studies~\cite{PhysRevLett.133.176601,davenport2025excitonberryology} highlight the role of symmetry classification in elucidating the interplay between crystal symmetries and excitonic topology. While these studies provide valuable insights at specific high-symmetry points, a systematic symmetry-based classification of the eigenstates of the BSE across the Brillouin zone within an \textit{ab initio} framework is still largely unexplored. 

In addition to providing insight and understanding, symmetries can be used to reduce the computational cost of the calculations. In the context of one-electron states, most widely used electronic structure codes, such as \textsc{Vasp}~\cite{PhysRevB.54.11169}, \textsc{QuantumEspresso}~\cite{Giannozzi_2009}, and \textsc{Abinit}~\cite{GONZE20092582}, routinely incorporate symmetry-based optimizations to reduce computational cost and ensure physically meaningful results. This is used by calculating the electronic states for momenta within the irreducible part of the BZ and using symmetry to transform them to states with momenta within the rest of the BZ. Furthermore, in the context of lattice dynamics, similar symmetry-based approaches are used. Codes such as \textsc{Phonopy}~\cite{phonopy} and \textsc{Phono3py}~\cite{doi:10.7566/JPSJ.92.012001} leverage crystal symmetries to reduce the cost of calculating dynamical matrices. However, similar ideas to use symmetry to reduce the computational costs in problems where excitons at finite c.m. are essential have not yet been employed. Many important physical phenomena such as exciton-phonon scattering~\cite{doi:10.1021/acs.nanolett.3c00732, doi:10.1021/acs.nanolett.4c01508, PhysRevLett.122.187401, PhysRevLett.125.107401, PhysRevB.105.085111, Ramanpaper}, indirect optical transitions~\cite{PhysRevLett.122.187401}, and exciton thermalization dynamics~\cite{PhysRevB.111.184305} require detailed knowledge of excitonic states for a dense c.m. momentum grid. Exciton-phonon coupling, in particular, determines linewidths and exciton thermalization  dynamics~\cite{PhysRevB.111.184305}. In systems like MoS$_2$~\cite{doi:10.1021/acs.nanolett.3c00732,PhysRevLett.111.216805}, achieving convergence of calculations of the optical spectra demands fine sampling of the c.m.  momenta in the BZ, especially near band extrema where small momentum shifts can strongly modulate coupling strength. Similarly, modeling exciton dynamics via the Boltzmann equation~\cite{PhysRevB.111.184305} and indirect optical spectra calculations in hBN, silicon, and bilayer MoS$_2$~\cite{PhysRevLett.108.167402,hbnoptical,PhysRevLett.122.187401,PhysRevLett.105.136805,PhysRev.111.1245,PhysRevLett.34.155} require extensive sampling of finite c.m.  momenta excitonic states. However, performing BSE calculations across such fine c.m. momenta meshes is computationally prohibitive for most materials. This further underscores the need for a symmetry-adapted approach to excitons that is grounded in an \textit{ab initio} framework.

In this work, we develop a comprehensive symmetry-based formalism for exciton calculations within an \textit{ab initio} framework. By applying space group operations on exciton wave functions at finite c.m. momentum, $\mathbf{Q}$, we reconstruct the wave functions for momenta in the full BZ from computations for $\mathbf{Q}$ restricted to the irreducible wedge. We further classify excitonic eigenstates into irreducible representations, providing a rigorous symmetry-resolved picture of exciton physics. At $\mathbf{Q} = 0$, we use symmetry-adapted bases to block diagonalize the BSE Hamiltonian, reducing both diagonalization time and memory usage.

By systematically incorporating crystal symmetry into excitonic theory, our approach delivers both conceptual and computational advances. It enables scalable BSE calculations for complex systems and provides a robust foundation for interpreting exciton phenomena through the lens of symmetry.

\section{\label{sec:level2}THEORETICAL FORMALISM}

\subsection{\label{sec:level2a}Preliminaries: Space group symmetries and time-reversal symmetry in one-electron wave functions}

In periodic solids, the translational symmetry of the lattice ensures that the one-electron Hamiltonian $\hat{\mathcal{H}}$ commutes with the lattice translation operator, \(\hat{T}_{\mathbf{R}}\), where \(\mathbf{R}\) is a Bravais lattice vector. As a result, $\hat{\mathcal{H}}$ can be written as a direct sum of independent Hamiltonians at each crystal momentum $\mathbf{k}$ in the Brillouin zone,  
\begin{equation}
\hat{\mathcal{H}} = \bigoplus_{\mathbf{k}} \hat{\mathcal{H}}_{\mathbf{k}}
\end{equation}
This block-diagonal structure implies that the wave functions, $\phi_{n,\mathbf{k}}(\mathbf{r})$, can be chosen to be eigenstates of the translation operators, leading to the Bloch form
\begin{equation}\label{eq:blochform}
\phi_{n,\mathbf{k}}(\mathbf{r}) = e^{i\mathbf{k}\cdot\mathbf{r}} \, u_{n,\mathbf{k}}(\mathbf{r})
\end{equation}
where $u_{n,\mathbf{k}}(\mathbf{r})$ is periodic with the lattice. The cell periodic part $u_{n,\mathbf{k}}(\mathbf{r})$ is commonly expanded in a plane-wave basis as  
\begin{equation}\label{eq:4}
u_{n,\mathbf{k}}(\mathbf{r}) = \sum_{\mathbf{G}} c_{n,\mathbf{k}} (\mathbf{G}) e^{i \mathbf{G} \cdot \mathbf{r}}
\end{equation}
where $\mathbf{G}$ are reciprocal lattice vectors and $c_{n,\mathbf{k}}(\mathbf{G})$ are the expansion coefficients. This representation naturally incorporates translational symmetry and facilitates the treatment of additional crystal symmetries.

The symmetry properties of Bloch wave functions are central to understanding electronic band structures and the selection rules governing optical transitions. Due to the space-group symmetry of the underlying crystal lattice, Bloch wave functions are constrained to transform in specific ways under the corresponding symmetry operations. These transformations determine the irreducible representations associated with the wave functions.

Let $\hat{P}_{\{\mathcal{R}_{\mathbf{t}} \,|\, \mathbf{t}\}}$ represent a symmetry operator associated with $\{\mathcal{R}_{\mathbf{t}} \,|\, \mathbf{t}\} \in \mathcal{G}$, where $\mathcal{R}_{\mathbf{t}}$ denotes a point-group operation (such as rotation, mirror reflection, or inversion), $\mathbf{t}$ is a fractional translation, and $\mathcal{G}$ is the crystal space group. Due to the underlying symmetry of the lattice, $\hat{P}_{\{\mathcal{R}_{\mathbf{t}} \,|\, \mathbf{t}\}}$ commutes with the Hamiltonian, \([\hat{P}_{\{\mathcal{R}_{\mathbf{t}} \,|\, \mathbf{t}\}}, \hat{\mathcal{H}}] = 0\). $\hat{P}_{\{\mathcal{R}_{\mathbf{t}} \,|\, \mathbf{t}\}}$ maps $\hat{\mathcal{H}}_{\mathbf{k}}$ to $\hat{\mathcal{H}}_{\mathcal{R}_{\mathbf{t}}\mathbf{k}}$ via $\hat{\mathcal{H}}_{\mathcal{R}_{\mathbf{t}}\mathbf{k}} = \hat{P}_{\{\mathcal{R}_{\mathbf{t}} \,|\, \mathbf{t}\}} \, \hat{\mathcal{H}}_{\mathbf{k}} \, \hat{P}_{\{\mathcal{R}_{\mathbf{t}} \,|\, \mathbf{t}\}}^{-1}$ (See Appendix A). This implies that if a wavevector $\mathbf{k}$ is mapped to $\mathcal{R}_{\mathbf{t}}\mathbf{k}$ by a symmetry operation, the energy eigenvalues corresponding to $\mathbf{k}$ and $\mathcal{R}_{\mathbf{t}}\mathbf{k}$ remain same {\em{i.e.}} 
\(\epsilon_{n, \mathcal{R}_{\mathbf{t}}\mathbf{k}} = \epsilon_{n, \mathbf{k}}\). The action of the symmetry operator on the Bloch wave function transforms it according to
\begin{equation}\label{eq:3}
\hat{P}_{\{\mathcal{R}_{\mathbf{t}} \,|\, \mathbf{t}\}} \, \phi_{n,\mathbf{k}}(\mathbf{r}) 
= \phi_{n,\mathbf{k}}\left( \mathcal{R}_{\mathbf{t}}^{-1} (\mathbf{r} - \mathbf{t}) \right)
\end{equation}
Utilizing Bloch's form as defined in Eq. \ref{eq:blochform}, the transformation becomes
\begin{equation}\label{eq:5}
\hat{P}_{\{{\mathcal{R}_{\mathbf{t}}}|{\mathbf{t}}\}} \phi_{n,\mathbf{k}}(\mathbf{r}) = u^{\mathbf{t}}_{n,{\mathcal{R}_{\mathbf{t}}}\mathbf{k}}(\mathbf{r}) e^{i{\mathcal{R}_{\mathbf{t}}}\mathbf{k}.(\mathbf{r}-{\mathbf{t}})}
\end{equation}
Here the cell-periodic function \(u^{\mathbf{t}}_{n,{\mathcal{R}_{\mathbf{t}}}\mathbf{k}}(\mathbf{r}) = u_{n,\mathbf{k}}(\mathcal{R}_{\mathbf{t}}^{-1}(\mathbf{r}-\mathbf{t})) \)  and can be written as
\begin{equation}\label{eq:6}
u^{\mathbf{t}}_{n,{\mathcal{R}_{\mathbf{t}}}\mathbf{k}}(\mathbf{r}) =  \sum_{\mathbf{G}}c_{n,\mathbf{k}} ({\mathcal{R}_{\mathbf{t}}^{-1}} \mathbf{G}) e^{-i \mathbf{G} \cdot {\mathbf{t}}}e^{i \mathbf{G} \cdot \mathbf{r}}
\end{equation}
Substituting this into Eq.~\ref{eq:5} yields the transformation rule for the plane-wave coefficients at \({\mathcal{R}_{\mathbf{t}}}\mathbf{k}\) in terms of those at \(\mathbf{k}\):
\begin{equation}\label{eq:7}
c^{\mathbf{t}}_{n,{\mathcal{R}_{\mathbf{t}}}\mathbf{k}} (\mathbf{G}) = c_{n,\mathbf{k}} ({\mathcal{R}_{\mathbf{t}}^{-1}} \mathbf{G}) e^{-i (\mathbf{G} + {\mathcal{R}_{\mathbf{t}}} \mathbf{k}) \cdot {\mathbf{t}}}
\end{equation}
This equation captures how the plane-wave components of the Bloch wave function transform under a space group operation. In systems without degeneracy and spin, this relation uniquely determines the coefficients at \({\mathcal{R}_{\mathbf{t}}} \mathbf{k}\), up to a common phase factor across all bands.

When spin degrees of freedom are included, the symmetry properties of Bloch wave functions must account for both spatial transformations and their induced effects on spinors. Although lattice symmetries act on spatial coordinates in \(\mathbb{R}^3\), they also act on the spin degrees of freedom, which transform under \(\text{SU}(2)\), the double cover of the spatial rotation group \(\text{SO}(3)\).

A spinor Bloch wave function can be expressed as a two-component column vector:
\begin{equation}\label{eq:8}
\Phi_{n,\mathbf{k}}(\mathbf{r}) =
\begin{bmatrix}
\phi_{n,\mathbf{k},\uparrow}(\mathbf{r}) \\
\phi_{n,\mathbf{k},\downarrow}(\mathbf{r})
\end{bmatrix}
=
\begin{bmatrix}
u_{n,\mathbf{k},\uparrow}(\mathbf{r}) e^{i\mathbf{k} \cdot \mathbf{r}} \\
u_{n,\mathbf{k},\downarrow}(\mathbf{r}) e^{i\mathbf{k} \cdot \mathbf{r}}
\end{bmatrix}
\end{equation}
where \(u_{n,\mathbf{k},\sigma}(\mathbf{r})\) are periodic functions and \(\sigma = \uparrow, \downarrow\) denotes the spin index.

A general symmetry operation involving spin can be written as the direct product of a spatial operation and a corresponding spinor transformation:
\begin{equation}\label{eq:9}
 \hat{P}^{sp}_{\{{\mathcal{R}_{\mathbf{t}}}|{\mathbf{t}}\}} = \hat{P}_{\{{\mathcal{R}_{\mathbf{t}}}|{\mathbf{t}}\}} \otimes \hat{\mathcal{T}}_{\mathcal{R}_{\mathbf{t}}}
\end{equation}
where \(\hat{P}^{sp}_{\{{\mathcal{R}_{\mathbf{t}}}|{\mathbf{t}}\}}\) denotes the full operator that corresponds to the symmetry operation \(\{{\mathcal{R}_{\mathbf{t}}}|{\mathbf{t}}\}\), and \(\hat{P}_{\{{\mathcal{R}_{\mathbf{t}}}|{\mathbf{t}}\}}\) and \(\hat{\mathcal{T}}_{\mathcal{R}_{\mathbf{t}}}\) denote its spatial and spin components, respectively. The dependence on \({\mathbf{t}}\) is dropped from \(\hat{\mathcal{T}}_{\mathcal{R}_{\mathbf{t}}}\) since fractional translations do not affect the spinor representation. Let \(\mathcal{T}^{\sigma,\sigma'}_{{\mathcal{R}_{\mathbf{t}}}}\) be the matrix elements of the \(\text{SU}(2)\) representation corresponding to \(\hat{\mathcal{T}}_{\mathcal{R}_{\mathbf{t}}}\). Since spinors transform under the \(\text{SU}(2)\) representation of rotations, their behavior is governed by the homomorphism between \(\text{SO}(3)\) and \(\text{SU}(2)\), given by
\begin{equation}\label{eq:10}
\mathcal{T} : \text{SO}(3) \to \text{SU}(2)
\end{equation}
This mapping is a two-to-one covering: each rotation \({\mathcal{R}_{\mathbf{t}}} \in \text{SO}(3)\) corresponds to two elements \(\pm \mathcal{T}_{{\mathcal{R}_{\mathbf{t}}}} \in \text{SU}(2)\). Consequently, under a full \(2\pi\) rotation, a spinor acquires a phase of \(-1\), reflecting the half-integer spin of electrons.

The action of the full symmetry operator \(\hat{P}^{sp}_{\{{\mathcal{R}_{\mathbf{t}}}|{\mathbf{t}}\}}\) on the spinor Bloch wave function \(\Phi_{n,\mathbf{k}}(\mathbf{r})\) is
\begin{equation}\label{eq:11}
 \hat{P}^{sp}_{\{{\mathcal{R}_{\mathbf{t}}}|{\mathbf{t}}\}}\Phi_{n,\mathbf{k}}(\mathbf{r}) = \mathcal{T}_{{\mathcal{R}_{\mathbf{t}}}}\Phi^{\mathbf{t}}_{n,{\mathcal{R}_{\mathbf{t}}\mathbf{k}}}(\mathbf{r})
\end{equation}
Here the spinor wave function, $\Phi^{\mathbf{t}}_{n,{\mathcal{R}_{\mathbf{t}}\mathbf{k}}}(\mathbf{r})$ is defined as $\begin{bmatrix}
\phi^{\mathbf{t}}_{n,\mathcal{R}_{\mathbf{t}}\mathbf{k},\uparrow}(\mathbf{r}) \\
\phi^{\mathbf{t}}_{n,\mathcal{R}_{\mathbf{t}}\mathbf{k},\downarrow}(\mathbf{r})
\end{bmatrix}$.
The explicit form of the \(\text{SU}(2)\) spinor matrix, $\mathcal{T}_{{\mathcal{R}_{\mathbf{t}}}}$ for a rotation by an angle, $\theta$, about the $\hat{n}$ axis is
\begin{equation}\label{eq:12}
\mathcal{T}_{{\mathcal{R}_{\mathbf{t}}}}=
\begin{bmatrix}
\cos \left(\frac{\theta}{2} \right) - i n_z \sin \left(\frac{\theta}{2} \right) & (-n_y - i n_x) \sin \left(\frac{\theta}{2} \right) \\
(n_y - i n_x) \sin \left(\frac{\theta}{2} \right) & \cos \left(\frac{\theta}{2} \right) + i n_z \sin \left(\frac{\theta}{2} \right)
\end{bmatrix}
\end{equation}
This unitary transformation guarantees that the spinors are transformed correctly under spatial rotations. For a general rotation, the directions \(\hat{n}=(n_x,n_y,n_z)\) and \(\theta\) are chosen based on \({\mathcal{R}_{\mathbf{t}}}\). Because \(\mathcal{T}_{{\mathcal{R}_{\mathbf{t}}}} \in \text{SU}(2)\), spinor wave functions obey a different transformation law from scalar wave functions. This leads to the emergence of half-integer representations and the necessity to use double groups with distinct irreducible representations.

A key consequence of the \(\text{SU}(2)\)-\(\text{SO}(3)\) homomorphism is the sign ambiguity in group multiplication. If two spatial rotations \(({\mathcal{R}_{\mathbf{t}}})_i\) and \(({\mathcal{R}_{\mathbf{t}}})_j\) combine to form \(({\mathcal{R}_{\mathbf{t}}})_k\), that is,
\begin{equation}\label{eq:13}
({\mathcal{R}_{\mathbf{t}}})_i ({\mathcal{R}_{\mathbf{t}}})_j = ({\mathcal{R}_{\mathbf{t}}})_k
\end{equation}
then their corresponding spin representations satisfy
\begin{equation}\label{eq:14}
(\mathcal{T}_{{\mathcal{R}_{\mathbf{t}}}})_i (\mathcal{T}_{{\mathcal{R}_{\mathbf{t}}}})_j = \pm (\mathcal{T}_{{\mathcal{R}_{\mathbf{t}}}})_k
\end{equation}
The additional sign arises from the double cover nature of \(\text{SU}(2)\) over \(\text{SO}(3)\). Specifically, a \(2\pi\) rotation changes the sign of a spinor wave function, a fundamental property underlying fermionic statistics, and the behavior of electrons in systems with spin-orbit coupling. This sign ambiguity is reflected in the group multiplication rules of double groups, where the symmetry elements remain the same, but the signs depend on the chosen branch of the \(\text{SU}(2)\) representation corresponding to a given \(\text{SO}(3)\) rotation.

So far, our discussion has been restricted to cases where the energy eigenstates are nondegenerate, both with and without spin-orbit coupling. However, the presence of degeneracies introduces an additional complexity in the symmetry analysis, since multiple wave functions may mix under the action of symmetry operations. Under a symmetry operation \({\mathcal{R}_{\mathbf{t}}}\), a Bloch state \(|\phi_{m\mathbf{k}}\rangle\) can transform into a linear combination of states within the degenerate manifold. This transformation can be written as
\begin{equation}\label{eq:15}
\hat{P}_{\{{\mathcal{R}_{\mathbf{t}}}|{\mathbf{t}}\}} |m,\mathbf{k} \rangle = \sum_n \mathcal{D}^{n,m}_{{\mathcal{R}_{\mathbf{t}}}\mathbf{k}}(\{{\mathcal{R}_{\mathbf{t}}}|{\mathbf{t}}\}) |n, {\mathcal{R}_{\mathbf{t}}}\mathbf{k} \rangle
\end{equation}
From this point onward, we use the simplified notation \(|n,\mathbf{k} \rangle\) to represent one-particle Bloch states. The coefficients \(\mathcal{D}^{n,m}_{{\mathcal{R}_{\mathbf{t}}}\mathbf{k}}(\{{\mathcal{R}_{\mathbf{t}}}|{\mathbf{t}}\})\) define a unitary transformation matrix associated with the symmetry operation \(\{{\mathcal{R}_{\mathbf{t}}}|{\mathbf{t}}\}\) within the degenerate subspace. These elements are obtained from the overlap between the rotated wave functions and the original states:
\begin{equation}\label{eq:16}
\mathcal{D}^{n,m}_{{\mathcal{R}_{\mathbf{t}}}\mathbf{k}}(\{{\mathcal{R}_{\mathbf{t}}}|{\mathbf{t}}\}) = \langle n, {\mathcal{R}_{\mathbf{t}}}\mathbf{k} | \hat{P}_{\{{\mathcal{R}_{\mathbf{t}}}|{\mathbf{t}}\}} | m,\mathbf{k} \rangle
\end{equation}
A particularly important case arises when \(\{{\mathcal{R}_{\mathbf{t}}}|{\mathbf{t}}\} \in \mathcal{G}_{\mathbf{k}}\), where \(\mathcal{G}_{\mathbf{k}}\) denotes the little group of the wave vector \(\mathbf{k}\), i.e., the subset of the space group symmetry operations that leave \(\mathbf{k}\) invariant modulo a reciprocal lattice vector \(\mathbf{G}\), such that \({\mathcal{R}_{\mathbf{t}}}\mathbf{k} = \mathbf{k} \pm \mathbf{G}\). In this case, the transformation reduces to
\begin{equation}\label{eq:17}
\mathcal{D}^{n,m}_{{\mathcal{R}_{\mathbf{t}}}\mathbf{k}}(\{{\mathcal{R}_{\mathbf{t}}}|{\mathbf{t}}\}) = \langle n,\mathbf{k} | \hat{P}_{\{{\mathcal{R}_{\mathbf{t}}}|{\mathbf{t}}\}}| m,\mathbf{k} \rangle = U^{n,m}_{\mathbf{k}}(\{{\mathcal{R}_{\mathbf{t}}}|{\mathbf{t}}\}),
\end{equation}
where \(U^{n,m}_{\mathbf{k}}(\{{\mathcal{R}_{\mathbf{t}}}|{\mathbf{t}}\})\) corresponds to an irreducible representation of the symmetry \(\{{\mathcal{R}_{\mathbf{t}}}|{\mathbf{t}}\}\) within the little group \(\mathcal{G}_{\mathbf{k}}\). This result implies that, in the presence of degeneracies, symmetry operations act within the degenerate subspace according to irreducible representations of the little group.

For nondegenerate states, the symmetry representation simplifies to a one-dimensional character \(e^{i\theta}\), reflecting the fact that the wave function transforms into itself up to a complex phase under \(\{{\mathcal{R}_{\mathbf{t}}}|{\mathbf{t}}\}\). In contrast, for an \(n\)-fold degenerate manifold, arising, for instance, due to spin, crystal symmetries, or fundamental symmetries such as time reversal, the representation becomes \(n\)-dimensional. These higher-dimensional irreducible representations determine the structure of degeneracies in the band structure and restrict the allowed symmetry-adapted basis states.

In the case of nonspinor wave functions (i.e., in the absence of spin-orbit coupling), the irreducible representation of the symmetry operation \(\{{\mathcal{R}_{\mathbf{t}}}|{\mathbf{t}}\}\) takes the form
\begin{align}\label{eq:18}
U^{m,n}_{\mathbf{k}}(\{{\mathcal{R}_{\mathbf{t}}}|{\mathbf{t}}\}) &= 
\sum_{\mathbf{G}} c^{*}_{m,\mathbf{k}} ({\mathcal{R}_{\mathbf{t}}} \mathbf{k} - \mathbf{k} + {\mathcal{R}_{\mathbf{t}}}\mathbf{G}) c_{n,\mathbf{k}} (\mathbf{G}) \notag \\
& \times e^{-i({\mathcal{R}_{\mathbf{t}}}\mathbf{k} + {\mathcal{R}_{\mathbf{t}}}\mathbf{G}) \cdot {\mathbf{t}}}
\end{align}
When spin-orbit coupling is taken into account, the Bloch wave functions become spinors. The transformation properties must then include the spin rotation induced by the symmetry operation. In this case, the representation generalizes to:
\begin{equation}\label{eq:spinorirrep}
U^{m,n}_{\mathbf{k}}(\{{\mathcal{R}_{\mathbf{t}}}|{\mathbf{t}}\}) = 
\sum_{\sigma \sigma'} \mathcal{T}^{\sigma \sigma'}_{{\mathcal{R}_{\mathbf{t}}}}  \langle m,\mathbf{k},\sigma | \hat{P}_{\{{\mathcal{R}_{\mathbf{t}}}|{\mathbf{t}}\}}| n,\mathbf{k},\sigma' \rangle
\end{equation}
where \(\mathcal{T}^{\sigma \sigma'}_{{\mathcal{R}_{\mathbf{t}}}}\) denotes the matrix elements of the spinor representation corresponding to the point group operation \({\mathcal{R}_{\mathbf{t}}}\). This formulation captures the combined effect of spatial and spin rotations on the symmetry behavior of Bloch spinors.

We have discussed the action of space group symmetries on Bloch wave functions, where spatial symmetry operations act through unitary transformations representing the real-space rotations, reflections, and translations, along with their corresponding action in reciprocal space. We now turn to the role of time-reversal symmetry, which differs fundamentally from space group operations due to its anti-unitary nature.

Time-reversal symmetry, unlike space group operations, reverses both the momentum and spin of a system and involves complex conjugation. In the spinless case, the time-reversal symmetry operator, \(\hat{P}_{\Theta}\), reduces to the complex conjugation operator \(\hat{\mathcal{C}}\) and relates Bloch states at \(\mathbf{k}\) and \(-\mathbf{k}\) through:
\begin{equation}\label{eq:19}
\hat{P}_{\Theta}\ket{n,\mathbf{k}} = \ket{n,-\mathbf{k}}
\end{equation}
In position representation, using the identities \(\hat{\mathcal{C}}^{\dagger}\mathbf{r}\hat{\mathcal{C}} = \mathbf{r}\) and \(\hat{\mathcal{C}}^{\dagger}\mathbf{k}\hat{\mathcal{C}} = -\mathbf{k}\), this becomes:
\begin{equation}\label{eq:19}
\hat{P}_{\Theta}\phi_{n,\mathbf{k}}(\mathbf{r}) = \phi_{n,-\mathbf{k}}(\mathbf{r}) = \phi^{*}_{n,\mathbf{k}}(\mathbf{r})
\end{equation}
In the plane-wave basis, this implies the relation:
\begin{equation}\label{eq:TRCoeff}
c_{n,-\mathbf{k}}(\mathbf{G}) = c^{*}_{n,\mathbf{k}}(-\mathbf{G})
\end{equation}
This relation allows wave functions at \(-\mathbf{k}\) points in the BZ to be constructed from their time-reversal partners, \(\mathbf{k}\), where they have been calculated explicitly. However, the \(\mathbf{\Gamma}\), \textit{i.e.} \(\mathbf{k} = 0\), is a special point, as at this point \(\mathbf{k} = -\mathbf{k}\). While the numerically obtained wave functions carry an arbitrary diagonalization phase, $e^{i\alpha_{n,\mathbf{k}}}$ at any \(\mathbf{k}\) point for a band index $n$, this phase causes Eq. \ref{eq:TRCoeff} to not be automatically followed at the \(\Gamma\) point. In order to restore time-reversal symmetry at \(\Gamma\), one can compute the following representation for each band:
\begin{equation}\label{eq:nospTR}
\Theta_{n,\mathbf{\Gamma}} = \bra{n,\mathbf{\Gamma}}\left(\hat{P}_{\Theta}\ket{n,\mathbf{\Gamma}}\right) = e^{-2i\alpha_{n,\mathbf{\Gamma}}}
\end{equation}
The resulting quantity \(\Theta_{n,\mathbf{\Gamma}}\) is a phase which can be used to construct the time reversal symmetric wave function as \(\ket{n,\mathbf{\Gamma}} ^{TR}= \sqrt{\Theta_{n,\mathbf{\Gamma}}}\ket{n,\mathbf{\Gamma}}\). Hereafter, the action of time-reversal symmetry will be denoted without explicitly writing the brackets 
on the right, with the understanding that it represents the operation itself.

For systems with spin-orbit coupling, time-reversal symmetry acts on spinor Bloch wave functions through the operator \(\hat{P}^{sp}_{\Theta} = -i \sigma_y \hat{\mathcal{C}}\), where \(\sigma_y\) is the Pauli matrix acting in spin space. Its action in position representation on the spinor wave function is:
\begin{align}\label{eq:19}
\hat{P}^{sp}_{\Theta}\Phi_{n,\mathbf{k}}(\mathbf{r}) 
 &=
\begin{bmatrix}
\phi_{n,-\mathbf{k},\uparrow}(\mathbf{r}) \\
\phi_{n,-\mathbf{k},\downarrow}(\mathbf{r})
\end{bmatrix}
=-i\sigma_y\hat{\mathcal{C}}\begin{bmatrix}
\phi_{n,\mathbf{k},\uparrow}(\mathbf{r}) \\
\phi_{n,\mathbf{k},\downarrow}(\mathbf{r})
\end{bmatrix} \notag\\
&=
\begin{bmatrix}
-\phi^{*}_{n,\mathbf{k},\downarrow}(\mathbf{r}) \\
\phi^{*}_{n,\mathbf{k},\uparrow}(\mathbf{r})
\end{bmatrix}
\end{align}
In the plane-wave basis, this leads to the following conditions:
\begin{equation}\label{eq:19}
c_{n,-\mathbf{k},\uparrow}(\mathbf{G}) = -c^{*}_{n,\mathbf{k},\downarrow}(\mathbf{-G}) \quad ; \quad
c_{n,-\mathbf{k},\downarrow}(\mathbf{G}) = c^{*}_{n,\mathbf{k},\uparrow}(\mathbf{-G})
\end{equation}
To enforce this condition at the \(\Gamma\) point, one computes the time-reversal representation for spinor states analogous to the spinless case, as described in Eq. \ref{eq:nospTR}. These symmetry constraints are essential for ensuring the correct transformation behavior of wave functions under time-reversal, particularly in systems with spin-orbit coupling and in the construction of time reversal symmetric excitonic states, which will be discussed next.

\subsection{\label{sec:level2b}Excitons and translational symmetry}
The study of electron–hole excitations from the many-body ground state \(|N,0\rangle\) to an excited state \(|N,S\rangle\) can be rigorously formulated within the framework of the two-particle Green's function and its equation of motion, the Bethe–Salpeter equation (BSE)~\cite{PhysRevLett.49.1519}. Here, \(S\) labels the excitation, while \(N\) denotes the conserved total number of electrons. Such excitations correspond to the creation of an electron in a conduction-band state and the removal of an electron from a valence-band state (equivalently, the creation of a hole).

Following the  work of Strinati~\cite{PhysRevLett.49.1519}, the electron–hole amplitude is defined from the electron–hole correlation function as
\begin{equation}\label{eq:ehamp}
    \Psi_S(\mathbf{x}, \mathbf{x}') = -\langle N, 0 | \hat{\Phi}^\dagger(\mathbf{x}') \hat{\Phi}(\mathbf{x}) | N, S \rangle
\end{equation}
where \(\hat{\Phi}^\dagger(\mathbf{x}')\) and \(\hat{\Phi}(\mathbf{x})\) create and annihilate an electron at positions \(\mathbf{x}'\) and \(\mathbf{x}\), respectively.

Within the Tamm–Dancoff approximation (TDA), \(\Psi_S(\mathbf{x}, \mathbf{x}')\) admits the expansion
\begin{equation}\label{eq:ehamp_tda}
\Psi_S(\mathbf{x}, \mathbf{x}') = \sum_v^{\text{occ}} \sum_c^{\text{empty}} A^S_{vc} \, \Phi_v^*(\mathbf{x}') \, \Phi_c(\mathbf{x})
\end{equation}
where \(\Phi_v(\mathbf{x}')\) and \(\Phi_c(\mathbf{x})\) are single-particle valence and conduction wave functions. The expansion coefficients are given by
\begin{equation}
A^S_{vc} \;=\; \langle N,0 \,|\, \hat{b}_c \,\hat{a}_v \,|\, N,S \rangle
\end{equation}
with \(\hat{a}_v^\dagger\) creating a hole in state \(v\) and \(\hat{b}_c^\dagger\) creating an electron in state \(c\).

In periodic systems, the BSE Hamiltonian respects lattice translational symmetry. As a result, the two-particle translational operator \(\hat{T}^{\text{ex}}_{\mathbf{R}}\) commutes with the BSE Hamiltonian \(\hat{\mathcal{H}}_{\mathrm{BSE}}\) (see Appendix B). This allows the excitations to be labeled by a conserved total momentum \(\mathbf{Q}\), which is a good quantum number. The electron-hole amplitude associated with the finite momentum \(\mathbf{Q}\) can then be written as:
\begin{equation}\label{eq:excwavefn}
\Psi_{S,\mathbf{Q}}(\mathbf{r}_e, \mathbf{r}_h) = \sum_{v,c,\mathbf{k}} A^{S,\mathbf{Q}}_{v,c,\mathbf{k}}\, \Phi_{c,\mathbf{k}}(\mathbf{r}_e)\, \Phi^{*}_{v,\mathbf{k}-\mathbf{Q}}(\mathbf{r}_h)
\end{equation}
where \(\Phi_{c,\mathbf{k}}(\mathbf{r}_e)\) and \(\Phi_{v,\mathbf{k}-\mathbf{Q}}(\mathbf{r}_h)\) are Bloch wave functions evaluated at the electron and hole coordinates \(\mathbf{r}_e\) and \(\mathbf{r}_h\), respectively. 
From this point onward, we use the ket, \(|\, S,\mathbf{Q} \rangle\), to represent the electron-hole amplitude state whose spatial representation is given in Eq. (\ref{eq:excwavefn}).
Since the hole is generated by removing an electron at \(\mathbf{k}-\mathbf{Q}\), its momentum is \(-(\mathbf{k}-\mathbf{Q}) = \mathbf{Q}-\mathbf{k}\). This allows it to be represented by the time reversal of the valence-band electron state \(|v,\mathbf{k}-\mathbf{Q}\rangle\). The excited state can thus be expressed in the product basis as
\begin{equation}\label{eq:21}
|S,\mathbf{Q}\rangle = \sum_{v,c,\mathbf{k}} A_{v,c,\mathbf{k}}^{S,\mathbf{Q}}\, |v,\mathbf{k}-\mathbf{Q};\, c,\mathbf{k}\rangle
\end{equation}
with
\begin{equation}\label{eq:25}
|v,\mathbf{k}-\mathbf{Q};\, c,\mathbf{k}\rangle = \hat{P}_{\Theta}^{h}\, |v,\mathbf{k}-\mathbf{Q}\rangle \otimes |c,\mathbf{k}\rangle
\end{equation}
The excitonic Hilbert space is as a result a direct product of the electron and hole Hilbert spaces. 
Introducing center-of-mass and relative coordinates, \(\mathbf{R} = \alpha\mathbf{r}_e + \beta\mathbf{r}_h\) (with $\alpha+\beta=1$) and \(\mathbf{r} = \mathbf{r}_e - \mathbf{r}_h\),  respectively, Eq.~\eqref{eq:excwavefn} takes the Bloch-periodic form
\begin{equation}\label{eq:27}
\Psi_{S,\mathbf{Q}}(\mathbf{R}, \mathbf{r}) = e^{i\mathbf{Q} \cdot \mathbf{R}} F_{S,\mathbf{Q}}(\mathbf{R}, \mathbf{r})
\end{equation}
where the phase factor encodes the exciton’s total momentum and \(F_{S,\mathbf{Q}}(\mathbf{R}, \mathbf{r})\) contains the cell-periodic structure (see Appendix~B). This generalizes the single-particle Bloch theorem to the interacting two-particle excitonic case.
The coefficients \(A^{S,\mathbf{Q}}_{v,c,\mathbf{k}}\) and energies \(\Omega_{S,\mathbf{Q}}\) are obtained by solving the BSE eigenvalue problem at fixed \(\mathbf{Q}\):
\begin{align}\label{eq:BSEQ}
&\left( \epsilon_{c,\mathbf{k}} - \epsilon_{v,\mathbf{k}-\mathbf{Q}} \right) A^{S,\mathbf{Q}}_{v,c,\mathbf{k}}
+ \sum_{v', c', \mathbf{k}'} \langle v, \mathbf{k} - \mathbf{Q}; c, \mathbf{k} | K^{eh}
\notag\\
&| v', \mathbf{k}' - \mathbf{Q}; c', \mathbf{k}' \rangle \, A^{S,\mathbf{Q}}_{v',c',\mathbf{k}'}
= \Omega_{S,\mathbf{Q}} A^{S,\mathbf{Q}}_{v,c,\mathbf{k}}
\end{align}

\subsection{\label{sec:level2c}Excitons and time-reversal symmetry}

We begin by defining the time-reversal operator for excitons as the tensor product of the time-reversal operators acting individually on the valence hole and conduction electron:
\begin{equation}\label{eq:26}
\hat{P}_{\Theta}^{ex} = \hat{P}_{\Theta}^{h} \otimes \hat{P}_{\Theta}^{e}
\end{equation}
Since the excitonic Hamiltonian commutes with the time-reversal operator, the exciton eigenstates at momentum \(\mathbf{Q}\) and \(-\mathbf{Q}\) are related by time reversal symmetry (see Appendix C):
\begin{equation}\label{eq:timerevexc}
\Omega_{S, \mathbf{Q}} = \Omega_{S, -\mathbf{Q}}, \quad \hat{P}_{\Theta}^{ex} \ket{S,\mathbf{Q}} = \ket{S,-\mathbf{Q}}
\end{equation}
In the real-space representation, this relation resembles the transformation of single-particle wave functions under time-reversal and can be expressed as:
\begin{equation}\label{eq:26}
\hat{P}_{\Theta}^{ex} \Psi_{S,\mathbf{Q}}(\mathbf{r}_e,\mathbf{r}_h) = \Psi^{*}_{S,\mathbf{Q}}(\mathbf{r}_e,\mathbf{r}_h) = \Psi_{S,-\mathbf{Q}}(\mathbf{r}_e,\mathbf{r}_h)
\end{equation}
We now derive the explicit transformation of the exciton expansion coefficients under time-reversal symmetry by examining the action of \(\hat{P}_{\Theta}^{ex}\) on the excitonic state. The transformation takes the form:
\begin{align}\label{eq:timeexc}
|S, -\mathbf{Q}\rangle &= \hat{P}^{ex}_{\Theta} |S, \mathbf{Q}\rangle \notag \\
&= \sum_{v,c,\mathbf{k}} (A^{S,\mathbf{Q}}_{v,c,\mathbf{k}} )^{*}
\left[\hat{P}^{h}_{\Theta}\hat{P}^{h}_{\Theta} |v, \mathbf{k} - \mathbf{Q}\rangle\right] 
\otimes 
\left[\hat{P}^{e}_{\Theta} |c, \mathbf{k}\rangle\right]
\end{align}
As \(\hat{P}_{\Theta}^{ex}\) is an antilinear operator, it 
not only transforms the basis states but also complex conjugates the exciton coefficients.

Using the known time-reversal properties of the one-electron Bloch states, namely
\begin{equation}
\hat{P}_{\Theta}^{h} |v, \mathbf{k} - \mathbf{Q}\rangle = |v, -\mathbf{k} + \mathbf{Q} \rangle, \quad \hat{P}_{\Theta}^{e} |c, \mathbf{k}\rangle = |c, -\mathbf{k} \rangle
\end{equation}
Eq.~\ref{eq:timeexc} becomes
\begin{align}\label{eq:timeexc2}
|S, -\mathbf{Q}\rangle &= \sum_{v,c,\mathbf{k}} (A^{S,\mathbf{Q}}_{v,c,\mathbf{k}} )^{*}
\left[\hat{P}^{h}_{\Theta}|v, -\mathbf{k} + \mathbf{Q} \rangle\right] \otimes |c, -\mathbf{k} \rangle \notag \\
&= \sum_{v,c,\mathbf{k}} (A^{S,\mathbf{Q}}_{v,c,-\mathbf{k}} )^{*}
\left[\hat{P}^{h}_{\Theta}|v, \mathbf{k} + \mathbf{Q} \rangle\right] \otimes |c, \mathbf{k} \rangle
\end{align}
By comparing this with the exciton state \(|S, -\mathbf{Q}\rangle\) expressed in the product-state basis \(\hat{P}^{h}_{\Theta}|v, \mathbf{k} + \mathbf{Q} \rangle \otimes |c, \mathbf{k} \rangle\), we identify the transformation law for the exciton wave function coefficients under time reversal:
\begin{equation}\label{eq:exccoeffTR}
\tilde{A}^{S,-\mathbf{Q}}_{v,c,\mathbf{k}} = (A^{S,\mathbf{Q}}_{v,c,-\mathbf{k}} )^{*}
\end{equation}
This result holds provided that the one-electron wave functions are constructed on a time-reversal symmetric \(\mathbf{k}\) grid and are generated over the full Brillouin zone, ensuring time-reversal symmetry is preserved in both spinless and spinor systems. In particular, at \(\mathbf{k} = 0\), the wave functions must explicitly satisfy time-reversal symmetry (see subsection A). This typically requires removing the arbitrary diagonalization phases at \(\Gamma\), which can be achieved by computing the time-reversal representation matrices \(\Theta_{n,\Gamma}\), as discussed in Eq. \ref{eq:nospTR}.

\subsection{\label{sec:level2d}Excitons and space group  symmetries}
We define the action of a space group symmetry operation on the excitonic state as the tensor product of symmetry operators acting on the valence hole and conduction electron:
\begin{equation}\label{eq:28}
\hat{P}^{ex}_{\{{\mathcal{R}_{\mathbf{t}}}|{\mathbf{t}}\}} = \hat{P}^{h}_{\{{\mathcal{R}_{\mathbf{t}}}|{\mathbf{t}}\}} \otimes \hat{P}^{e}_{\{{\mathcal{R}_{\mathbf{t}}}|{\mathbf{t}}\}}
\end{equation}
Since the Bethe-Salpeter Hamiltonian commutes with the excitonic symmetry operator, i.e., \([\hat{P}^{ex}_{\{{\mathcal{R}_{\mathbf{t}}}|{\mathbf{t}}\}}, \hat{\mathcal{H}}_{\mathrm{BSE}}] = 0\), and the electron-hole amplitudes exhibit Bloch periodicity under symmetry actions (see Appendix D), we obtain:
\begin{equation}\label{eq:excrot}
\Omega_{S, {\mathcal{R}_{\mathbf{t}}}\mathbf{Q}} = \Omega_{S, \mathbf{Q}}, \quad 
|{S, {\mathcal{R}_{\mathbf{t}}}\mathbf{Q}}\rangle =  \hat{P}^{ex}_{\{{\mathcal{R}_{\mathbf{t}}}|{\mathbf{t}}\}}|S, \mathbf{Q}\rangle
\end{equation}
This symmetry relation guarantees that excitonic eigenstates transform consistently under lattice symmetries, analogous to single-particle Bloch states.

Utilizing this framework, we can derive a relation between the coefficients \(A^{S,\mathbf{Q}}_{v,c,\mathbf{k}}\) and those of the transformed state \(\tilde{A}^{S,{\mathcal{R}_{\mathbf{t}}}\mathbf{Q}}_{v,c,\mathbf{k}}\), akin to the transformation properties of single-particle coefficients \(c_{n, \mathbf{k}}(\mathbf{G})\) and \(c^{{\mathbf{t}}}_{n, {\mathcal{R}_{\mathbf{t}}}\mathbf{k}}(\mathbf{G})\) as seen in Eq.~\ref{eq:7}.

The transformation of the excitonic state under symmetry becomes:
\begin{align}\label{eq:30}
&|S, {\mathcal{R}_{\mathbf{t}}}\mathbf{Q}\rangle =  \hat{P}^{ex}_{\{{\mathcal{R}_{\mathbf{t}}}|{\mathbf{t}}\}}|S, \mathbf{Q}\rangle = 
\notag \\
&\sum_{v,c,\mathbf{k}} A^{S,\mathbf{Q}}_{v,c,\mathbf{k}} 
\left[\hat{P}^{h}_{\{{\mathcal{R}_{\mathbf{t}}}|{\mathbf{t}}\}}\hat{P}_{\Theta}^{h}|v, \mathbf{k} - \mathbf{Q}\rangle\right] 
\otimes 
\left[\hat{P}^{e}_{\{{\mathcal{R}_{\mathbf{t}}}|{\mathbf{t}}\}}|c, \mathbf{k}\rangle\right]
\end{align}
and, expanding the action of these operators in their respective degenerate subspaces, we obtain
\begin{align}\label{eq:31}
|S, {\mathcal{R}_{\mathbf{t}}}\mathbf{Q}\rangle = \sum_{v',c',\mathbf{k}} 
\bigg[
\sum_{v,c} A^{S,\mathbf{Q}}_{v,c,{\mathcal{R}_{\mathbf{t}}^{-1}}\mathbf{k}} 
\mathcal{L}^{v',v}_{\mathbf{k}-{\mathcal{R}_{\mathbf{t}}}\mathbf{Q}}(\{{\mathcal{R}_{\mathbf{t}}}|{\mathbf{t}}\}) 
\notag \\
\otimes 
\mathcal{D}^{c',c}_{\mathbf{k}}(\{{\mathcal{R}_{\mathbf{t}}}|{\mathbf{t}}\})
\bigg] |v', \mathbf{k} - {\mathcal{R}_{\mathbf{t}}}\mathbf{Q}\rangle \otimes |c', \mathbf{k}\rangle
\end{align}
where the matrix elements for conduction and valence band transformations are defined as
\begin{equation}\label{eq:32}
\mathcal{D}^{c',c}_{\mathbf{k}}(\{{\mathcal{R}_{\mathbf{t}}}|{\mathbf{t}}\}) =
\langle c',\mathbf{k}|\hat{P}^{e}_{\{{\mathcal{R}_{\mathbf{t}}}|{\mathbf{t}}\}}|c,{\mathcal{R}_{\mathbf{t}}}^{-1}\mathbf{k}\rangle
\end{equation}
\begin{align}\label{eq:33}
&\mathcal{L}^{v',v}_{\mathbf{k}-{\mathcal{R}_{\mathbf{t}}}\mathbf{Q}}(\{{\mathcal{R}_{\mathbf{t}}}|{\mathbf{t}}\}) =
\bra{ v',\mathbf{k}-{\mathcal{R}_{\mathbf{t}}}\mathbf{Q}}{\hat{P}_{\Theta}^{h}\dagger} \hat{P}^{h}_{\{{\mathcal{R}_{\mathbf{t}}}|{\mathbf{t}}\}}
\notag \\
& \times \hat{P}_{\Theta}^{h}\ket{
v,{\mathcal{R}_{\mathbf{t}}^{-1}}(\mathbf{k}-{\mathcal{R}_{\mathbf{t}}}\mathbf{Q})}= \left[\mathcal{D}^{v',v}_{\mathbf{k}-{\mathcal{R}_{\mathbf{t}}}\mathbf{Q}}(\{{\mathcal{R}_{\mathbf{t}}}|{\mathbf{t}}\})\right]^{*}
\end{align}
The transformed exciton coefficients are then given by
\begin{align}\label{eq:34}
\tilde{A}^{S, {\mathcal{R}_{\mathbf{t}}}\mathbf{Q}}_{v,c,\mathbf{k}} &= 
\sum_{v',c'} A^{S,\mathbf{Q}}_{v',c',{\mathcal{R}_{\mathbf{t}}}^{-1}\mathbf{k}} \notag \\
& \left[\mathcal{D}^{v,v'}_{\mathbf{k}-{\mathcal{R}_{\mathbf{t}}}\mathbf{Q}}(\{{\mathcal{R}_{\mathbf{t}}}|{\mathbf{t}}\})\right]^{*} 
\otimes \mathcal{D}^{c,c'}_{\mathbf{k}}(\{{\mathcal{R}_{\mathbf{t}}}|{\mathbf{t}}\})
\end{align}
This can be written compactly using matrix-vector multiplication, where the transformation matrix has elements
\begin{equation}\label{eq:35}
\mathcal{M}_{v,c;v',c'}^{\mathbf{k},\mathbf{Q}}(\{{\mathcal{R}_{\mathbf{t}}}|{\mathbf{t}}\}) = 
[\mathcal{D}^{v,v'}_{\mathbf{k}-{\mathcal{R}_{\mathbf{t}}}\mathbf{Q}}(\{{\mathcal{R}_{\mathbf{t}}}|{\mathbf{t}}\})]^* 
\otimes 
\mathcal{D}^{c,c'}_{\mathbf{k}}(\{{\mathcal{R}_{\mathbf{t}}}|{\mathbf{t}}\})
\end{equation}
and the matrix form becomes
\begin{align}\label{eq:36}
[\tilde{A}^{S, {\mathcal{R}_{\mathbf{t}}}\mathbf{Q}}_{\mathbf{k}}]_{v,c} &= 
\sum_{v',c'} \mathcal{M}_{v,c;v',c'}^{\mathbf{k},\mathbf{Q}}(\{{\mathcal{R}_{\mathbf{t}}}|{\mathbf{t}}\}) 
A^{S,\mathbf{Q}}_{v',c',{\mathcal{R}_{\mathbf{t}}}^{-1}\mathbf{k}} \notag \\
&= \left[\mathcal{M}^{\mathbf{k},\mathbf{Q}}(\{{\mathcal{R}_{\mathbf{t}}}|{\mathbf{t}}\}) \cdot A^{S,\mathbf{Q}}_{{\mathcal{R}_{\mathbf{t}}}^{-1}\mathbf{k}}\right]_{v,c}
\end{align}
Thus the transformation of the exciton coefficient vector at each \(\mathbf{k}\)-point simplifies to
\begin{equation}\label{eq:37}
\tilde{A}^{S, {\mathcal{R}_{\mathbf{t}}}\mathbf{Q}}_{\mathbf{k}} =
\mathcal{M}^{\mathbf{k},\mathbf{Q}}(\{{\mathcal{R}_{\mathbf{t}}}|{\mathbf{t}}\}) \cdot A^{S, \mathbf{Q}}_{{\mathcal{R}_{\mathbf{t}}}^{-1}\mathbf{k}}
\end{equation}
When dealing with spinor wave functions, the symmetry operator \(\hat{P}^{h/e}_{\{{\mathcal{R}_{\mathbf{t}}}|{\mathbf{t}}\}}\) must be replaced with the product \(\hat{P}^{h/e}_{\{{\mathcal{R}_{\mathbf{t}}}|{\mathbf{t}}\}} \otimes \hat{\mathcal{T}}^{v/c}_{\mathcal{R}_{\mathbf{t}}}\), where \(\hat{\mathcal{T}}_{\mathcal{R}_{\mathbf{t}}}\) denotes the SU(2) rotation corresponding to the SO(3) spatial symmetry \({\mathcal{R}_{\mathbf{t}}}\). This ensures that spinor structure is correctly accounted for, consistent with the spin representations corresponding to Eq.~\ref{eq:14}.

\subsection{\label{sec:level2e}Symmetry classification of excitons}

Exciton bands can be classified according to the irreducible representations of the symmetry group, in analogy with single-particle bands. This classification is possible now that we have established how excitonic states transform under symmetry operations. The definition of the little group is also analogous, with the key distinction being that, for excitons, it is defined with respect to the center-of-mass (c.m.) momentum \( \mathbf{Q} \). The little group \( \mathcal{G}_{\mathbf{Q}} \) consists of all symmetry operations \( \{{\mathcal{R}_{\mathbf{t}}}|{\mathbf{t}}\} \) that leave \( \mathbf{Q} \) invariant up to a reciprocal lattice vector, i.e., \( {\mathcal{R}_{\mathbf{t}}} \mathbf{Q} = \mathbf{Q} \pm \mathbf{G} \). For any such \( \{{\mathcal{R}_{\mathbf{t}}}|{\mathbf{t}}\} \in \mathcal{G}_{\mathbf{Q}} \), the excitonic state \( |S, \mathbf{Q} \rangle \) transforms as
\begin{equation}\label{eq:38}
 \hat{P}^{ex}_{\{{\mathcal{R}_{\mathbf{t}}}|{\mathbf{t}}\}}|S,\mathbf{Q}\rangle = \sum_{S'=1}^{N_{exc}} \mathcal{K}^{S',S}_{\mathbf{Q}}(\{{\mathcal{R}_{\mathbf{t}}}|{\mathbf{t}}\}) |S',\mathbf{Q}\rangle
\end{equation}
Here, \( \mathcal{K}_{\mathbf{Q}}(\{{\mathcal{R}_{\mathbf{t}}}|{\mathbf{t}}\}) \) forms a representation of the little group \( \mathcal{G}_{\mathbf{Q}} \) within the invariant subspace spanned by the excitonic states \( \{|S, \mathbf{Q}\rangle\}_{N_{exc}} \). Its matrix elements are given by
\begin{equation}\label{eq:39}
\mathcal{K}^{S',S}_{\mathbf{Q}}(\{{\mathcal{R}_{\mathbf{t}}}|{\mathbf{t}}\}) = \langle S',\mathbf{Q}| \hat{P}^{ex}_{\{{\mathcal{R}_{\mathbf{t}}}|{\mathbf{t}}\}}| S,\mathbf{Q} \rangle
\end{equation}
The action of \( \hat{P}^{ex}_{\{{\mathcal{R}_{\mathbf{t}}}|{\mathbf{t}}\}} \) on \( |S,\mathbf{Q}\rangle \), using Eqs.~\ref{eq:30} and \ref{eq:37}, leads to
\begin{equation}\label{eq:40}
 \hat{P}^{ex}_{\{{\mathcal{R}_{\mathbf{t}}}|{\mathbf{t}}\}}|S,\mathbf{Q}\rangle = \sum_{v',c',\mathbf{k}'}\Tilde{A}_{v',c',\mathbf{k}'}^{S,\mathbf{Q}}[\hat{P}_{\Theta}^{h}|v',\mathbf{k'-Q}\rangle]\otimes|c',\mathbf{k'}\rangle
\end{equation}
To compute the matrix elements of the representation, we expand \( |S', \mathbf{Q} \rangle \) in the electron-hole basis and evaluate the overlap:
\begin{align}\label{eq:41}
\mathcal{K}^{S',S}_{\mathbf{Q}}(\{{\mathcal{R}_{\mathbf{t}}}|{\mathbf{t}}\}) &= \sum_{v,c,v',c',\mathbf{k},\mathbf{k'}} (A_{v,c,\mathbf{k}}^{S',\mathbf{Q}})^{*}\tilde{A}_{v', c', \mathbf{k'}}^{S, \mathbf{Q}} 
\notag \\ 
&\langle v, \mathbf{k} - \mathbf{Q} |{\hat{P}_{\Theta}^{h\dagger}}\hat{P}_{\Theta}^{h}|v', \mathbf{k'} - \mathbf{Q} \rangle \langle c, \mathbf{k} | c', \mathbf{k'} \rangle
\end{align}
Using orthonormality of the electronic states and \( {\hat{P}_{\Theta}^{h,\dagger}} \hat{P}_{\Theta}^{h} = I \), this expression simplifies to
\begin{equation}\label{eq:42}
 \mathcal{K}^{S',S}_{\mathbf{Q}}(\{{\mathcal{R}_{\mathbf{t}}}|{\mathbf{t}}\}) = \sum_{v, c, v', c', \mathbf{k}, \mathbf{k'}} (A_{v,c,\mathbf{k}}^{S',\mathbf{Q}})^{*} \tilde{A}_{v', c', \mathbf{k'}}^{S, \mathbf{Q}} \delta_{v,v'} \delta_{c,c'} \delta_{\mathbf{k},\mathbf{k'}}
\end{equation}
yielding
\begin{equation}\label{eq:43}
 \mathcal{K}^{S',S}_{\mathbf{Q}}(\{{\mathcal{R}_{\mathbf{t}}}|{\mathbf{t}}\}) = \sum_{v, c,\mathbf{k}} (A_{v,c,\mathbf{k}}^{S',\mathbf{Q}} )^{*}\tilde{A}_{v, c, \mathbf{k}}^{S, \mathbf{Q}} 
\end{equation}
Finally, using Eq.~\ref{eq:37}, we obtain a compact form:
\begin{align}\label{eq:irrepexciton}
&\mathcal{K}^{S',S}_{\mathbf{Q}}(\{{\mathcal{R}_{\mathbf{t}}}|{\mathbf{t}}\}) \notag \\
&= \sum_{v,c,\mathbf{k}} \left[A_{\mathbf{k}}^{S', \mathbf{Q}}\right]_{v,c}^{*}\left[\mathcal{M}^{\mathbf{k},\mathbf{Q}}(\{{\mathcal{R}_{\mathbf{t}}}|{\mathbf{t}}\})A^{S,\mathbf{Q}}_{{\mathcal{R}_{\mathbf{t}}}^{-1}\mathbf{k}}\right]_{v,c}
\end{align}
Each representation \( \mathcal{K}_{\mathbf{Q}} \) is characterized by its trace, known as the character:
\begin{equation}\label{eq:45}
\mathcal{X}_{\mathcal{K}_{\mathbf{Q}}}(\{{\mathcal{R}_{\mathbf{t}}}|{\mathbf{t}}\}) = \mathrm{Tr}\big[\mathcal{K}_{\mathbf{Q}}(\{{\mathcal{R}_{\mathbf{t}}}|{\mathbf{t}}\})\big] = \sum_{S} \mathcal{K}_{\mathbf{Q}}^{S,S}(\{{\mathcal{R}_{\mathbf{t}}}|{\mathbf{t}}\})
\end{equation}
In general, the invariant subspace spanned by \( \{|S,\mathbf{Q}\rangle\}_{N_{exc}} \) may decompose into a direct sum of multiple irreducible representations. Therefore, the representation \( \mathcal{K}_{\mathbf{Q}} \) may be reducible and can be expressed as
\begin{equation}\label{eq:46}
\mathcal{K}_{\mathbf{Q}} = \oplus_{n} m_{\mathbf{Q}}^{\xi_{n}}\mathcal{K}_{\mathbf{Q}}^{\xi_{n}} 
\end{equation}
Here, $\mathcal{K}_{\mathbf{Q}}^{\xi_{n}}$ is the $n^{\text{th}}$ irreducible representation of $\mathcal{G}_{\mathbf{Q}}$ ($n = 1, \dots, N_{\xi}$), with $N_{\xi}$ denoting the total number of such representations. The coefficient $m_{\mathbf{Q}}^{\xi_n}$ indicates the multiplicity of $\mathcal{K}_{\mathbf{Q}}^{\xi_{n}}$ in $\mathcal{K}_{\mathbf{Q}}$.
 This multiplicity can be computed using
\begin{equation}\label{eq:characterformula}
m_{\mathbf{Q}}^{\xi_n} = \frac{1}{N_{\mathcal{G}_{\mathbf{Q}}}}\sum_{\{{\mathcal{R}_{\mathbf{t}}}|{\mathbf{t}}\}\in\mathcal{G}_{\mathbf{Q}}}\mathcal{X}_{\mathcal{K}_{\mathbf{Q}}}^{*}(\{{\mathcal{R}_{\mathbf{t}}}|{\mathbf{t}}\})\mathcal{X}_{\xi_n}(\{{\mathcal{R}_{\mathbf{t}}}|{\mathbf{t}}\})
\end{equation}
where \( N_{\mathcal{G}_{\mathbf{Q}}} \) is the order of the group and \( \mathcal{X}_{\xi_n}(\{{\mathcal{R}_{\mathbf{t}}}|{\mathbf{t}}\}) \) is the character of the group element $\{{\mathcal{R}_{\mathbf{t}}}|{\mathbf{t}}\}$ in the $n^{\text{th}}$ irreducible representation, which can be obtained from symmetry tools such as \textsc{spgrep}.

To assign each excitonic state to a specific irreducible representation, we use the corresponding projection operator:
\begin{equation}\label{eq:48}
    \hat{\mathcal{V}}^{(\xi_n)}_{ij;\mathbf{Q}} = \frac{d_{\xi_n}}{N_{\mathcal{G}_{\mathbf{Q}}}} \sum_{\{{\mathcal{R}_{\mathbf{t}}}|{\mathbf{t}}\} \in\mathcal{G}_{\mathbf{Q}} } \big[\Delta^{(\xi_n)}_{ij}(\{{\mathcal{R}_{\mathbf{t}}}|{\mathbf{t}}\})\big]^{*}\hat{P}^{ex}_{\{{\mathcal{R}_{\mathbf{t}}}|{\mathbf{t}}\}}
\end{equation}
Here, $ \hat{\mathcal{V}}^{(\xi_n)}_{ij;\mathbf{Q}} $ projects onto the subspace transforming as the irreducible representation $\xi_n$ with dimension $ d_{\xi_n} $ and $ \Delta^{(\xi_n)}_{ij}(\{{\mathcal{R}_{\mathbf{t}}}|{\mathbf{t}}\}) $ is the corresponding matrix representation of the symmetry operation $\{{\mathcal{R}_{\mathbf{t}}}|{\mathbf{t}}\}$.

Symmetry-based classification of excitons provides significant physical insight. States transforming under one-dimensional irreducible representations are nondegenerate, whereas higher-dimensional irreducible representations give rise to degeneracies. These degeneracies can be lifted by perturbations such as spin-orbit coupling or external fields. Moreover, the symmetry of excitonic states plays a crucial role in determining optical selection rules and polarization properties.

\subsection{\label{sec:level2f}Symmetry-adapted reduction in the BSE Hamiltonian}

For space group operations \(\{{\mathcal{R}_{\mathbf{t}}}|{\mathbf{t}}\}\in\mathcal{G}\), the excitonic symmetry operator commutes with the Bethe-Salpeter Hamiltonian, i.e., \([\hat{P}^{ex}_{\{{\mathcal{R}_{\mathbf{t}}}|{\mathbf{t}}\}}, \hat{\mathcal{H}}_{\mathrm{BSE}}] = 0\). This invariance under symmetry operations implies (by arguments similar to those presented in Appendix A for Bloch states) that the Hamiltonian blocks at exciton center-of-mass momenta \(\mathbf{Q}\) and \({\mathcal{R}_{\mathbf{t}}}\mathbf{Q}\) are related as

\begin{equation}\label{eq:49}
        \hat{P}_{\{{\mathcal{R}_{\mathbf{t}}}|{\mathbf{t}}\}}^{ex} \, \hat{\mathcal{H}}_\mathbf{Q}^{\text{BSE}} \, (\hat{P}^{ex}_{\{{\mathcal{R}_{\mathbf{t}}}|{\mathbf{t}}\}})^{-1} =
        \hat{\mathcal{H}}_{{\mathcal{R}_{\mathbf{t}}} \mathbf{Q}}^{\text{BSE}}
\end{equation}
This means that, for operations not in the little group \(\mathcal{G}_{\mathbf{Q}}\), the symmetry connects different momentum sectors. However, when \(\{{\mathcal{R}_{\mathbf{t}}}|{\mathbf{t}}\}\in\mathcal{G}_{\mathbf{Q}}\), i.e., they leave \(\mathbf{Q}\) invariant, the Hamiltonian satisfies

\begin{equation}\label{eq:50}
        \hat{P}_{\{{\mathcal{R}_{\mathbf{t}}}|{\mathbf{t}}\}}^{ex} \, \hat{\mathcal{H}}_\mathbf{Q}^{\text{BSE}} \, (\hat{P}^{ex}_{\{{\mathcal{R}_{\mathbf{t}}}|{\mathbf{t}}\}})^{-1} =
        \hat{\mathcal{H}}_{\mathbf{Q}}^{\text{BSE}}
\end{equation}
Thus \(\hat{\mathcal{H}}_{\mathbf{Q}}^{\text{BSE}}\) transforms as a representation of the little group \(\mathcal{G}_{\mathbf{Q}}\). In cases where this representation is reducible, as commonly happens at high-symmetry points like \(\mathbf{Q}=0\) or others with less symmetries, the Hamiltonian can be block diagonalized into subspaces associated with irreducible representations, greatly simplifying the diagonalization.

We now outline the general formalism for constructing symmetry-adapted irreducible blocks of the Hamiltonian, valid at both zero and finite \(\mathbf{Q}\). A key step involves using the projection operators [defined in Eq. \ref{eq:48}], which were earlier used for exciton classification. Here, they serve to build symmetry-adapted product state bases that isolate irreducible subspaces of the Hilbert space, enabling block diagonalization.

By applying these projectors to the exciton product basis defined in Eq. \ref{eq:25}, and using the transformation law in Eq. \ref{eq:15}, we obtain the symmetry-adapted basis as follows. For each irreducible representation $\xi_n$, we construct $d_{\xi_n}^2 N_c N_v N_{\mathbf{k}}$ linear combinations with $i, j = 1, \dots, d_{\xi_n}$ using Eq.~\ref{eq:51} given below:
\begin{align}\label{eq:51}
    &\ket{{\psi}^{(\xi_{n})}_{(i,j);v,c,\mathbf{k},\mathbf{Q}}} = \hat{\mathcal{V}}^{(\xi_{n})}_{ij;\mathbf{Q}}|v,\mathbf{k}-\mathbf{Q};\, c,\mathbf{k}\rangle
    \notag \\&= \frac{d_{\xi_{n}}}{N_{\mathcal{G}_{\mathbf{Q}}}} \sum_{\{{\mathcal{R}_{\mathbf{t}}}|{\mathbf{t}}\} \in\mathcal{G}_{\mathbf{Q}} } \big[\Delta^{(\xi_{n})}_{ij}(\{{\mathcal{R}_{\mathbf{t}}}|{\mathbf{t}}\})\big]^{*}\hat{P}^{ex}_{\{{\mathcal{R}_{\mathbf{t}}}|{\mathbf{t}}\}}|v,\mathbf{k}-\mathbf{Q};\, c,\mathbf{k}\rangle
    \notag\\ & =\frac{d_{\xi_{n}}}{N_{\mathcal{G}_{\mathbf{Q}}}} \sum_{\substack{v',c',\\
    \{{\mathcal{R}_{\mathbf{t}}}|{\mathbf{t}}\} \in\mathcal{G}_{\mathbf{Q}}} } \big[\Delta^{(\xi_{n})}_{ij}(\{{\mathcal{R}_{\mathbf{t}}}|{\mathbf{t}}\})\big]^{*}\left[\mathcal{D}^{v',v}_{{\mathcal{R}_{\mathbf{t}}}\mathbf{k}-\mathbf{Q}}(\{{\mathcal{R}_{\mathbf{t}}}|{\mathbf{t}}\})\right]^{*} 
    \notag \\ &  \otimes \mathcal{D}^{c',c}_{{\mathcal{R}_{\mathbf{t}}}\mathbf{k}}(\{{\mathcal{R}_{\mathbf{t}}}|{\mathbf{t}}\})|v',{\mathcal{R}_{\mathbf{t}}}\mathbf{k}-\mathbf{Q};\, c',{\mathcal{R}_{\mathbf{t}}}\mathbf{k}\rangle
\end{align}
However, these linear combinations form a set containing null as well as linearly dependent vectors. Using \textsc{Spgrep}, we extract a linearly independent subset $\{ \ket{{\psi}^{(\xi_n)}_{(i,j);v,c,\mathbf{k},\mathbf{Q}}} \}$ by removing the redundant and null vectors. This subset constitutes the symmetry-adapted basis corresponding to the irreducible representation $\xi_n$. The number of vectors in this set are $l_{\mathbf{Q}}^{\xi_n}=d_{\xi_n}m_{\mathbf{Q}}^{\xi_n}$, the  dimension times the multiplicity of the irreducible representation $\xi_n$ in the representation $\mathcal{K}_{\mathbf{Q}}$. By the orthogonality theorem, the bases constructed in this way for distinct irreducible representations are mutually orthogonal. Since the dimensions of the irreducible representations satisfy the relation $\sum_{n} d_{\xi_n}^2 = N_{\mathcal{G}_{\mathbf{Q}}}$, the expected number of symmetry-adapted basis vectors  $l_{\mathbf{Q}}^{\xi_n}$ associated with a given irreducible representation $\xi_n$ is approximately $d_{\xi_n}^2 N_c N_v N_{\mathbf{k}} / N_{\mathcal{G}_{\mathbf{Q}}}$, where the approximation arises from the finite size of the basis. In all the cases we have studied (see Section IV on results and discussion), we find that $l_{\mathbf{Q}}^{\xi_n}$ is close to this estimate.

When the BSE Hamiltonian is expressed in this basis, it takes a block-diagonal form:
\begin{equation}\label{eq:52}
    \begin{bmatrix}
        \mathcal{H}^{(1)} & 0 & 0 & \cdots & 0 \\
        0 & \mathcal{H}^{(2)} & 0 & \cdots & 0 \\
        0 & 0 & \mathcal{H}^{(3)} & \cdots & 0 \\
        \vdots & \vdots & \vdots & \ddots & \vdots \\
        0 & 0 & 0 & \cdots & \mathcal{H}^{({N_\xi})}
    \end{bmatrix}
\end{equation}
Each block $\mathcal{H}^{(n)}$ represents an independent sector corresponding to the irreducible representation $\xi_n$ of \(\mathcal{G}_{\mathbf{Q}}\).

This decomposition stems directly from group representation theory, where the exciton basis and symmetry operators form a reducible representation. Projection operators as discussed above allow one to isolate irreducible sectors, ensuring orthogonality and eliminating couplings between different symmetry blocks. 

Physically, these blocks correspond to excitonic states categorized by their symmetry. This classification is useful for identifying bright and dark excitons depending on their symmetry behavior under optical transitions. Computationally, this structure enables solving several smaller eigenvalue problems rather than a single large one, making the BSE calculations more tractable and symmetry-respecting.

\section{\label{sec:level3}COMPUTATIONAL DETAILS}

For our calculations, we used the experimental in-plane lattice constants for monolayer MoS$_2$ (3.168~\r{A}, S-S distance of 3.133~\r{A}). A vacuum spacing of 16~\r{A} was introduced along the out-of-plane direction to avoid spurious interactions between periodic images. 

Density functional theory (DFT) calculations were performed with the \textsc{Quantum Espresso}~\cite{Giannozzi_2009,Giannozzi_2017} package using the PBE generalized gradient approximation (GGA)~\cite{PhysRevLett.77.3865} for the exchange-correlation functional. The wave functions were expanded in plane waves up to an energy cutoff of 90 Ry. Spin-orbit coupling was explicitly included by using fully relativistic optimized norm-conserving Vanderbilt pseudopotentials~\cite{PhysRevB.88.085117} from the \textsc{PseudoDojo}~\cite{VANSETTEN201839} library. Self-consistent calculations~\cite{PhysRev.140.A1133} were performed on a $24 \times 24 \times 1$ \(\mathbf{k}\) grid, resulting in DFT band gap of 1661.9~meV for MoS$_2$.

Quasiparticle energies were obtained within the $\mathrm{G_0W_0}$~\cite{PhysRevB.34.5390} approximation using the \textsc{BerkeleyGW} package~\cite{DESLIPPE20121269,PhysRevB.62.4927,PhysRevLett.81.2312}, starting from the DFT wave functions and eigenvalues computed with \textsc{Quantum Espresso}. We employed the spinor implementation of \textsc{BerkeleyGW}~\cite{PhysRevB.106.115127}, wherein spin-orbit coupling is included non-perturbatively. The  dielectric function was evaluated using the generalized plasmon-pole model of Hybertsen and Louie~\cite{PhysRevB.34.5390}, with a $6 \times 6 \times 1$ $\mathbf{q}$ grid and 4000 occupied and unoccupied bands. Plane waves up to an energy cutoff of 25 Ry were used in the computation of dielectric function. The Brillouin-zone sampling was refined near $\mathbf{q} = 0$ using a nonuniform neck subsampling (NNS)~\cite{PhysRevB.95.035109} scheme with a fine nonuniform sampling of 10 points. Coulomb truncation was applied along the out-of-plane direction to eliminate interlayer interactions~\cite{PhysRevB.73.233103}. The resulting $\mathrm{GW}$ gap at the K point was 2553.9~meV for MoS$_2$.

The Bethe-Salpeter equation (BSE) was solved within the Tamm-Dancoff approximation using \textsc{BerkeleyGW}~\cite{DESLIPPE20121269,PhysRevB.106.115127,PhysRevB.62.4927,PhysRevLett.81.2312}. BSE calculations were performed for finite-$\mathbf{Q}$ points along the path $\Gamma$-M-K-$\Gamma$ of the Brillouin zone. The electron-hole interaction kernel and absorption calculations were done on a $24 \times 24 \times 1$ $\mathbf{k}$ grid with two valence and four conduction bands. The total product basis size was therefore 4608. The dielectric matrix was evaluated using plane waves up to the energy cutoff of 5 Ry in the BSE kernel calculations. One-electron wave functions at all the $\mathbf{k}$ points in the full Brillouin zone were constructed by rotating the wave functions generated in the irreducible Brillouin zone to preserve phase consistency at symmetry-related points. 

\section{\label{sec:level4}RESULTS AND DISCUSSION}
The formalism that we have developed in Sec. II is general. We use monolayer MoS$_2$ as a prototypical example to show the 
application of this formalism within an {\em ab initio} context. The nomenclature and labels used to represent the groups and their irreducible representations are adopted from Ref.~\cite{koster1963pointgroups}.
The crystal symmetry of monolayer MoS$_2$ is described by the point group $D_{3h}$. The little group at the center of the 
Brillouin zone ($\mathbf{k}=\Gamma$), 
$\mathcal{G}_{\Gamma}$, contains all the symmetry elements of the $D_{3h}$ group.
At other high-symmetry points in the Brillouin zone the little groups contain fewer symmetry elements---for example, at $\mathbf{k}=\mathrm{M}$ the little group, 
$\mathcal{G}_{\mathrm{M}}$, is $C_{2v}$, while at $\mathbf{k}=\mathrm{K}$, the little group, $\mathcal{G}_{\mathrm{K}}$, is $C_{3h}$.  
In the absence of spin-orbit coupling, the one-electron eigenstates are also eigenstates of the spin 
angular momentum operator. As a result, they can be labeled using the single-group irreducible representations of the little groups listed above. As our calculations include spin–orbit coupling, the one-electron eigenstates
are spinors as they are not eigenstates of the spin angular momentum operator. Consequently, the associated irreducible representations belong 
to the complex double groups, $D_{3h}^D$, $C_{2v}^D$, and $C_{3h}^D$ at the $\Gamma$, $\mathrm{M}$, and $\mathrm{K}$ points in the 
Brillouin zone, respectively. 
\begin{figure}
    \includegraphics[width=0.48\textwidth]{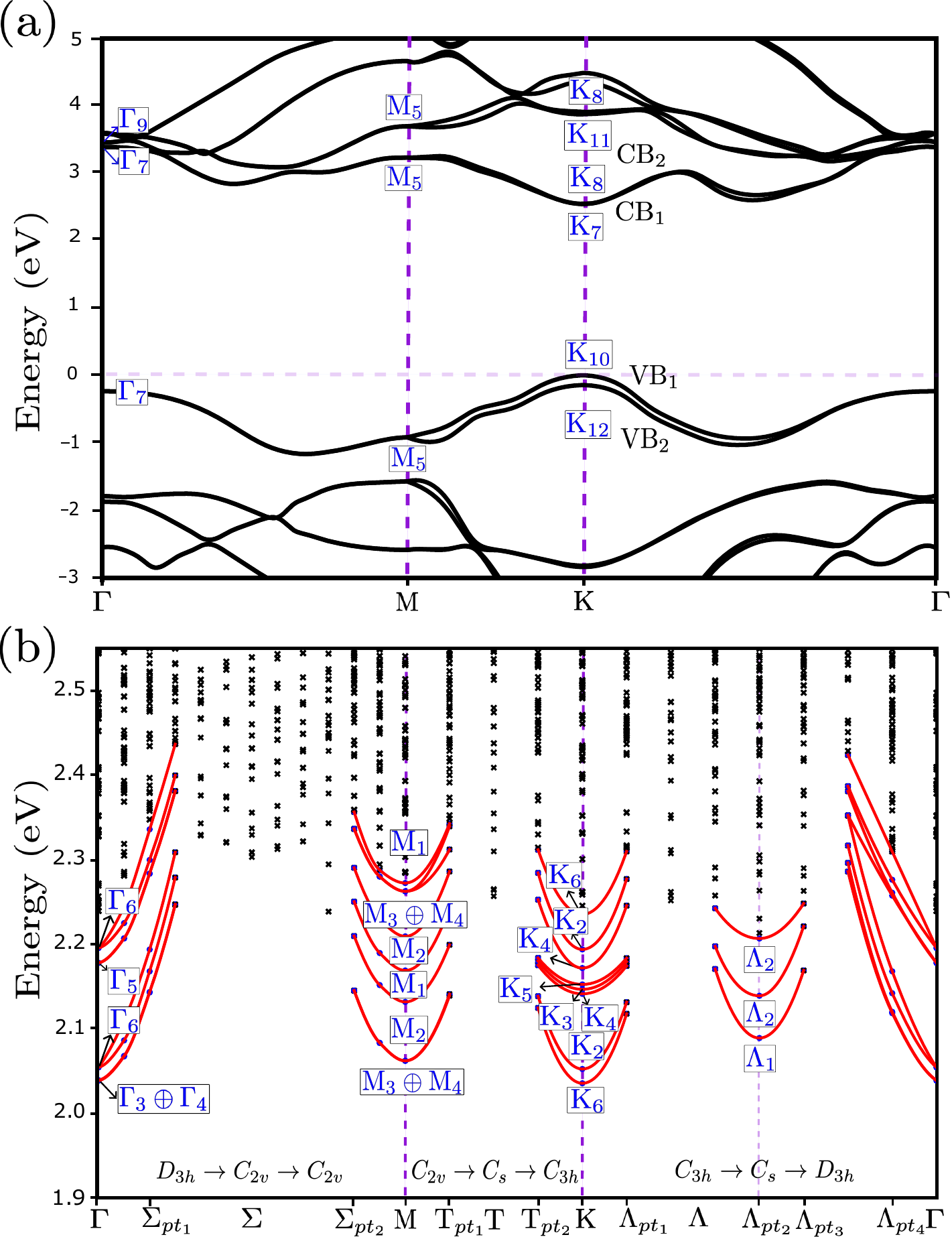}
    \caption{(a) GW electronic band structure of monolayer MoS$_{2}$ along the high-symmetry path $\Gamma$-M-K-$\Gamma$ in the Brillouin zone. The valence band maximum is set to $0$~eV. The double-group spinor irreducible representations associated with the bands are indicated at the high-symmetry points. (b) Exciton band structure of monolayer MoS$_{2}$ along the path $\Gamma$-M-K-$\Gamma$ in the Brillouin zone. The irreducible representations of the excitonic bands are labeled at the high-symmetry points $\Gamma$, M, and K. The irreducible representations at the labeled points along the high symmetry lines are tabulated in Table~\ref{tab:table1}. The evolution of the excitonic states at the transition between symmetry lines and high symmetry points illustrates the compatibility relations.}
    \label{fig:1}
\end{figure}
Figure~\ref{fig:1}(a) shows the quasiparticle band structure obtained using the $\mathrm{G}_{0}\mathrm{W}_{0}$ approximation to the self energy. The bands are plotted along the high-symmetry path $\Gamma-\mathrm{M}-\mathrm{K}-\Gamma$. We use the diagonal approximation of $\mathrm{G}_{0}\mathrm{W}_{0}$ to calculate the quasiparticle energies. Within this approximation, the quasiparticle wave functions are assumed to be the same as the corresponding DFT wave functions and the self-energy
operator only corrects the DFT eigenvalues to the corresponding quasiparticle energy. 
As a result, we use the DFT spinor wave functions to calculate the irreducible representations of the corresponding complex double groups at the high-symmetry points. Figure~\ref{fig:1}(a) also shows the
assignments of the irreducible representations at the high-symmetry points to the states that are closest to the band gap. 
For the first two valence and four conduction states at the $\Gamma$ point, the doubly degenerate states can be labeled by the 
irreducible representations $\Gamma_{7}$, $\Gamma_{7}$, and $\Gamma_{9}$ respectively. At the $\mathrm{M}$ point in the Brillouin 
zone, the labels are $\mathrm{M}_{5}$ for each pair of doubly degenerate states. At the $\mathrm{K}$ point, the one-dimensional 
representations of the valence bands are $\mathrm{K}_{12}$ and $\mathrm{K}_{10}$, respectively, and the representations are 
$\mathrm{K}_{7}$, $\mathrm{K}_{8}$, $\mathrm{K}_{11}$, and $\mathrm{K}_{8}$ for the spin–orbit split, nondegenerate states in the 
conduction band manifold. These irreducible representations are the same as those calculated from the \textsc{IrRep} 
package~\cite{IRAOLA2022108226}. 

In contrast to quasiparticles, excitons are composite bosons. The excitonic states are written as a linear combination of basis
states constructed from the tensor product of 
two fermionic states---electron and hole. In the absence of spin-orbit coupling, the total spin angular momentum of these basis states is given by the addition of the spin angular momenta of the constituent electron and hole states. 
This leads to the excitonic states being eigenstates of the total spin angular momentum operator. They are characterized by the eigenvalues of the square of the total spin operator, ${S}^{2}$, ($2\hbar^2$ for triplets and $\hbar^2$ for singlets), and $S_{z}$, the spin projection operator along the $z$ axis ($-1,0,1$ for triplets and $0$ for singlets).
In the presence of spin-orbit coupling, when the one-electron states are no longer eigenstates of the spin angular momentum operator,
the resulting excitons are also no longer eigenstates of the total spin angular momentum operator. Then, the excitonic 
eigenstates are linear combinations of singlet and triplet states. Nevertheless, in both cases (in the presence or absence of
spin orbit coupling), the symmetry classification of excitonic states is governed by the single-group irreducible representations 
of the little group at the center-of-mass momentum $\mathbf{Q}$ of the exciton.  Thus the relevant irreducible representations of the 
excitonic states are those of the single groups $D_{3h}$, $C_{2v}$, and $C_{3h}$ for $\mathbf{Q}=\Gamma$, $\mathbf{Q}=\mathrm{M}$ and 
$\mathbf{Q}=\mathrm{K}$, respectively.  

We implemented the formalism for applying spatial and time-reversal symmetries to excitonic states, as described in 
Secs.~\ref{sec:level2c} and ~\ref{sec:level2d}. In order to test the implementation, as a first step, we calculated the excitonic states at a point $\mathbf{Q}$ in the irreducible 
Brillouin zone. Upon rotating these states by a space group symmetry operation, $\{\mathcal{R}_{\mathbf{t}}|\mathbf{t}\}$, we obtained the excitonic states at 
the point $\mathcal{R}_{\mathbf{t}}\mathbf{Q}$. We compared these states to the corresponding excitonic states directly calculated at the 
rotated momentum $\mathcal{R}_{\mathbf{t}}\mathbf{Q}$. In an analogous manner, we compared the time-reversed exciton states at $\mathbf{Q}$ 
and $-\mathbf{Q}$. In both cases, we found exact agreement, up to an overall diagonalization phase, thereby confirming the 
correctness of our implementation of both space-group and time-reversal symmetries. 

Our implementation also allows for the direct computation of irreducible representations of the invariant subspaces within the 
exciton manifold at a given $\mathbf{Q}$ and their characters from the excitonic states, using Eqs.~\ref{eq:irrepexciton} and 
\ref{eq:characterformula} (Subsection~\ref{sec:level2f}). Using this approach, we obtained the irreducible representations of 
excitonic bands along the high-symmetry path $\Gamma-\mathrm{M}-\mathrm{K}-\Gamma$ in the Brillouin zone, as shown in 
Fig.~\ref{fig:1}(b). Consider the case $\mathbf{Q}=0$: the 1\textit{s}-like A excitons (A$_{1s}$) originate from the top valence band (VB$_1$) 
and the two lowest conduction bands (CB$_1$ and CB$_2$), near the $\mathrm{K}$/$\mathrm{-K}$ valleys~\cite{PhysRevLett.115.176801}. The complex double-group irreducible representations 
for VB$_1$, CB$_1$, and CB$_2$ are $\mathrm{K}_{10}$, $\mathrm{K}_{7}$, and $\mathrm{K}_{8}$ at the $\mathrm{K}$ valley (see Fig.~\ref{fig:1}), 
with conjugate irreducible representations $\mathrm{K}_{9}$, $\mathrm{K}_{8}$, and $\mathrm{K}_{7}$ at the $\mathrm{-K}$ valley.  
The transitions between CB$_1$ and VB$_1$ at the $\mathrm{K}$ and $\mathrm{-K}$ valley yield direct product states with irreducible representations given as  $\mathrm{K}_{7}^{*}\otimes \mathrm{K}_{10} = \mathrm{K}_3$ and $\mathrm{K}_{8}^{*}\otimes \mathrm{K}_{9} = \mathrm{K}_2$, respectively. If the exciton envelope function of $1s$ excitonic states transforms as $\mathrm{K}_{1}$, the resulting $1s$-like excitonic states correspond to $\mathrm{K}_3\oplus \mathrm{K}_2$. For the transitions involving CB$_2$ and VB$_1$ at the $\mathrm{K}$ and $\mathrm{-K}$ valley, we obtain the direct product states with irreducible representations as $\mathrm{K}_{8}^{*}\otimes \mathrm{K}_{10} = \mathrm{K}_4$ and $\mathrm{K}_{7}^{*}\otimes \mathrm{K}_{9} = \mathrm{K}_4$, respectively. As these transitions form the basis for excitons at the $\mathbf{Q}=\Gamma$ point in the excitonic band structure, we
use the compatibility relation for $C_{3h}\rightarrow D_{3h}$. This compatibility relation maps these irreducible representations
at $\mathrm{K}$ to irreducible representations at $\mathbf{Q}=\Gamma$ as $\mathrm{K}_3\oplus\mathrm{K}_2 \rightarrow \Gamma_{6}$ and $\mathrm{K}_4/\mathrm{K}_4 \rightarrow \Gamma_{3}/\Gamma_{4}$. Hence the first four A$_{1s}$ excitons transform as $\Gamma_{3}\oplus\Gamma_{4}\oplus\Gamma_{6}$. The classification obtained from our implementation is fully consistent with the physical and conceptual classification for A$_{1s}$ excitons (see Fig.~\ref{fig:1}b and Table.~\ref{tab:table1}).  

We next analyze the $1s$-like B excitons (B$_{1s}$), which originate from the valence band (VB$_2$) and the two lowest conduction bands (CB$_1$ and CB$_2$). The complex double-group irreducible representations for VB$_2$ at the $\mathrm{K}$ and $\mathrm{-K}$ valley is $\mathrm{K}_{10}$ and $\mathrm{K}_{11}$, respectively. The irreducible representations of the direct product of states corresponding to the transitions between CB$_1$ and VB$_2$ at $\mathrm{K}$ and $\mathrm{-K}$ valley are  $\mathrm{K}_{7}^{*}\otimes \mathrm{K}_{12} = \mathrm{K}_5$ and $\mathrm{K}_{8}^{*}\otimes \mathrm{K}_{11} = \mathrm{K}_6$, respectively. If the exciton envelope function of $1s$ transforms as $\mathrm{K}_{1}$, the corresponding direct product states are $\mathrm{K}_5\oplus \mathrm{K}_6$. For the transitions involving CB$_2$ and VB$_2$ at $\mathrm{K}$ and $\mathrm{-K}$ valley, we obtain $\mathrm{K}_{8}^{*}\otimes \mathrm{K}_{12} = \mathrm{K}_3$ and $\mathrm{K}_{7}^{*}\otimes \mathrm{K}_{11} = \mathrm{K}_2$, respectively, leading to the states belonging to $\mathrm{K}_3\oplus \mathrm{K}_{2}$. As discussed before, using the compatibility relation for  $C_{3h}\rightarrow D_{3h}$, this maps as $\mathrm{K}_5\oplus\mathrm{K}_6 \rightarrow \Gamma_{5}$ and $\mathrm{K}_3\oplus\mathrm{K}_2 \rightarrow \Gamma_{6}$. Therefore, the next four B$_{1s}$ excitons at the $\mathbf{Q}=\Gamma$ transform as $\Gamma_{5}\oplus\Gamma_{6}$. The classification obtained from our symmetry formalism is fully consistent with the physical and conceptual classification for B$_{1s}$ excitons, as well. (See Fig.~\ref{fig:1}b and Table.~\ref{tab:table1}).

Furthermore, we verified the validity of the irreducible representations at various finite center-of-mass momenta of excitons by explicitly tracking the compatibility relations between transitions of different symmetry groups at the high-symmetry points and the connecting symmetry lines. For $\Sigma_{pt_1}$ on the symmetry line $\Sigma$, the symmetry elements that form the group are $\{E,C_2,\sigma_h,\sigma_v\}$, which is isomorphic to the $C_{2v}$ group with symmetry elements $\{E,C_2,\sigma_v,\sigma_v'\}$ and the same character table. This group has four one-dimensional irreducible representations, denoted $\Sigma_{1}, \Sigma_{2}, \Sigma_{3}, \Sigma_{4}$. The compatibility relations from $D_{3h} \rightarrow C_{2v}$ are given by 
\[
\Gamma_3 \rightarrow \Sigma_{3}, \quad 
\Gamma_4 \rightarrow \Sigma_{4}, \quad 
\Gamma_6 \rightarrow \Sigma_{1} \oplus \Sigma_{2}, \quad 
\Gamma_5 \rightarrow \Sigma_3 \oplus \Sigma_4,
\]
which is in exact agreement with the independent symmetry classification obtained from our implementation. Higher-lying states are expected to follow the order predicted by these compatibility relations. However, due to band crossings and the presence of nearly degenerate states, the ordering of irreducible representations can change. This highlights the advantage of an explicit symmetry classification of excitonic bands, as it allows one to consistently identify states belonging to the same irreducible representations and to track them reliably, especially in the case of fine $\mathbf{k}$-point sampling where significant exciton overlap occurs between nearby points.  
\begin{figure}
    \includegraphics[width=0.475\textwidth]{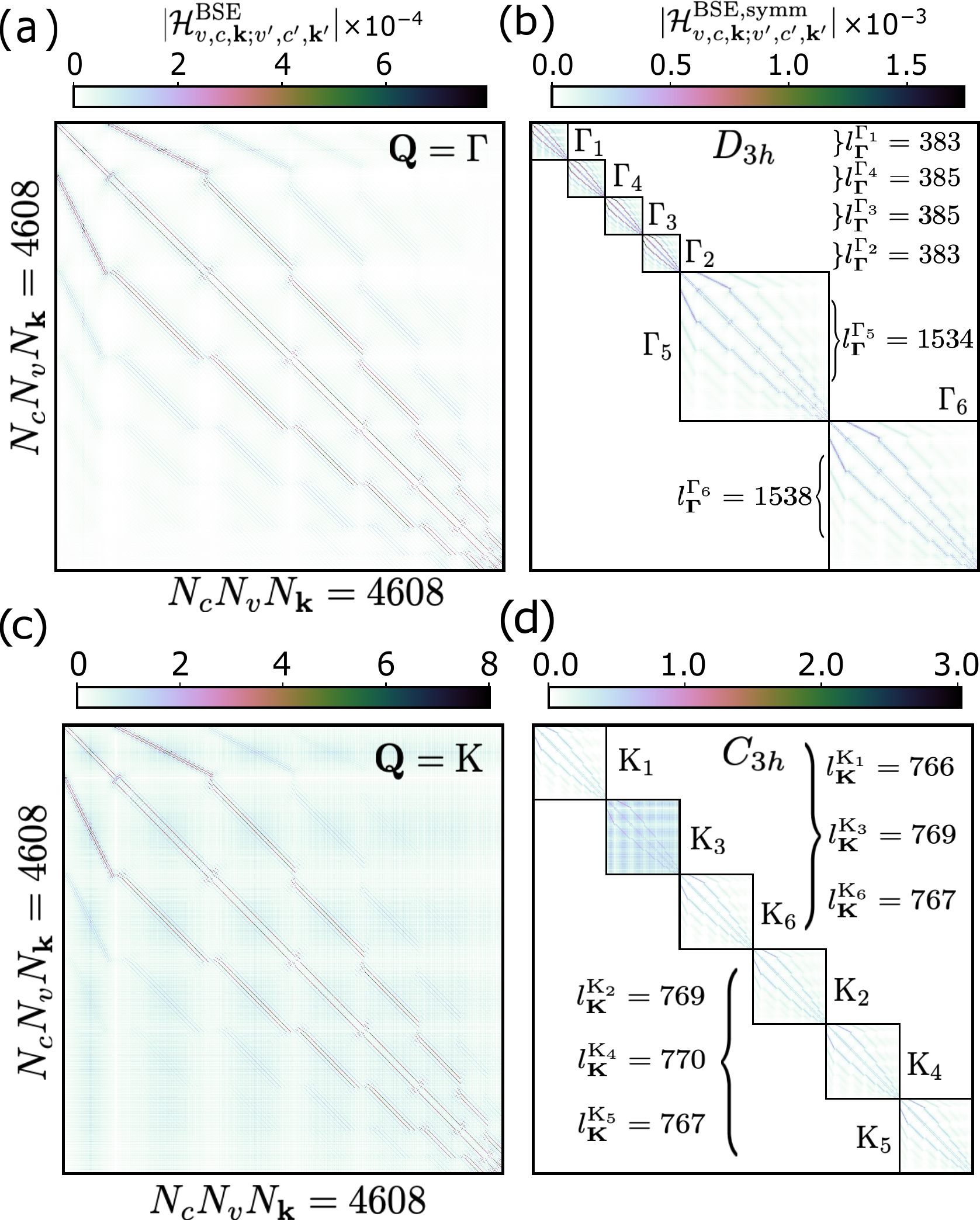}
    \caption{Panels (a) and (c) depict the full spinor BSE Hamiltonian constructed from two valence and four conduction bands on a $24 \times 24 \times 1$ $\mathbf{k}$-point grid, for exciton center-of-mass momenta $\mathbf{Q}=\Gamma$ and $\mathbf{Q}=\mathrm{K}$, respectively. Panels (b) and (d) show the corresponding block-diagonalized BSE Hamiltonians, resolved into blocks associated with the irreducible representations of the $D_{3h}$ and $C_{3h}$ symmetry groups at $\mathbf{Q}=\Gamma$ and $\mathbf{Q}=\mathrm{K}$, respectively. The dimensions of the blocks corresponding to each irreducible representation are indicated. The color bars represent the absolute values of the BSE Hamiltonian matrix elements for both the full and symmetry-adapted cases. For clarity, the diagonal matrix elements have been removed, and the color scale has been capped at a fixed maximum value to emphasize the block structure.}
    \label{fig:2}
\end{figure}
As one traverses the $\Sigma$ line toward $\Sigma_{pt_2}$, the ordering of the symmetry irreducible representations of the states changes. The irreducible representations obtained at $\Sigma_{pt_2}$ remain compatible with those at the $\mathrm{M}$ point, since the symmetry group is the same along the $\Sigma$ direction and $\mathbf{Q}=\mathrm{M}$. We then consider the $\mathrm{T}$ high-symmetry line from $\mathrm{M}$ to $\mathrm{K}$. Along this path, from the $\mathrm{M}$ point to $\mathrm{T}_{pt_1}$, the group reduces as $C_{2v} \rightarrow C_s = \{E,\sigma_h\}$. This group has two one-dimensional irreducible representations, with characters $1$ and $-1$ under $\sigma_h$, labeled as $\mathrm{T}_1$ and $\mathrm{T}_2$. Since the characters of $\sigma_h$ for $\mathrm{M}_3$ and $\mathrm{M}_4$ are $-1$, while those for $\mathrm{M}_1$ and $\mathrm{M}_2$ are $1$, the compatibility relations are
\[
\mathrm{M}_3/\mathrm{M}_4 \rightarrow \mathrm{T}_2, 
\quad \mathrm{M}_1/\mathrm{M}_2 \rightarrow \mathrm{T}_1.
\]
This relation holds for the first seven exciton states (see Table~\ref{tab:table1}), although the eighth and ninth states are accidentally degenerate in energy (i.e., not symmetry-protected). Consequently, the eighth exciton state at $\mathrm{M}$ is compatible with the ninth state at $\mathrm{T}_{pt_1}$. 

Following these connectivities, the irreducible representations at $\mathrm{T}_{pt_2}$ are shown in Table.~\ref{tab:table1}. One observes that some states shift within the manifold to preserve compatibility relations at the transition from $\mathrm{T}_{pt_2}$ to $\mathrm{K}$, where the symmetry changes from $C_{2v} \rightarrow C_{3h}$. The little group at $\mathbf{Q} = \mathrm{K}$ is $C_{3h}$, consisting of the symmetry elements $\{E, C_3, C_3^{-1}, \sigma_h, S_3, S_3^{-1}\}$. This group has six one-dimensional irreducible representations, labeled $\mathrm{K}_1$ through $\mathrm{K}_6$. The characters of $\sigma_h$ are $-1$ for $\mathrm{K}_4, \mathrm{K}_5, \mathrm{K}_6$, and $+1$ for $\mathrm{K}_1, \mathrm{K}_2, \mathrm{K}_3$. Thus, the compatibility relations between $\mathrm{T}_{pt_2}$ and $\mathrm{K}$ are
\[
\mathrm{K}_4,\mathrm{K}_5,\mathrm{K}_6 \rightarrow \mathrm{T}_2, 
\quad \mathrm{K}_1,\mathrm{K}_2,\mathrm{K}_3 \rightarrow \mathrm{T}_1.
\]
This correspondence is observed in our computed classifications, with the exception of the eighth state at $\mathrm{K}$, which was found to be compatible with the ninth state at $\mathrm{T}_{pt_2}$, again because of similar reasons discussed before. 

Along the $\Lambda$ line, one can similarly follow the compatibility relations from $C_{3h} \rightarrow C_s$. From $\Lambda_{pt_2}$ to $\Gamma$, \emph{i.e.}, from $C_{s} \rightarrow D_{3h}$, the characters of $\sigma_h$ are $-1, -1, 2$ for $\Gamma_3, \Gamma_4, \Gamma_6$, respectively, and $1, 1, -2$ for $\Gamma_1, \Gamma_2, \Gamma_5$, respectively. Therefore, the compatibility relations are
\begin{align*}
\Gamma_{3},\Gamma_{4} \rightarrow \Lambda_2, \quad 
\Gamma_{1},\Gamma_{2} \rightarrow \Lambda_1, \quad 
\\
\Gamma_{5} \rightarrow \Lambda_{2} \oplus \Lambda_2, \quad 
\Gamma_{6} \rightarrow \Lambda_{1} \oplus \Lambda_1.
\end{align*}
Our results are in excellent agreement with these predicted compatibility relations. 

In addition to proposing and implementing a formalism for symmetry-based classification of excitonic states, we employed projection operators (see subsection~\ref{sec:level2f}) to construct the symmetry-adapted linear combinations of the electron-hole direct-product-state basis for every irreducible representation of the little group $\mathcal{G}_{\mathbf{Q}}$. This procedure not only block diagonalizes the BSE Hamiltonian at that $\mathbf{Q}$ point into smaller blocks corresponding to distinct irreducible representations, but also provides an independent route for characterizing the excitons. While this approach naturally highlights the role of symmetry, in the present work instead of directly constructing the BSE kernel in the symmetry-adapted basis from the outset, we first compute the full kernel in the conventional electron–hole product basis and subsequently project it onto the symmetry-adapted basis. Consequently, the gain in computational efficiency in the current work arises mainly during the diagonalization step of the BSE Hamiltonian, where the matrix becomes block diagonal in the symmetry-adapted basis. We depict this reduction explicitly for $\mathbf{Q}=\Gamma$ and $\mathbf{Q}=\mathrm{K}$ in Fig.~\ref{fig:2}. For $\mathbf{Q} = \Gamma$, the little group has order $N_{\mathcal{G}_{\mathbf{Q}}} = 12$, with $d_{\xi_n} = 1$ for $\Gamma_1$, $\Gamma_2$, $\Gamma_3$, and $\Gamma_4$, and $d_{\xi_n} = 2$ for $\Gamma_5$ and $\Gamma_6$. This results in approximately 384 (= 4608/12) and 1536 (= 4 $\times$ 4608/12) symmetry-adapted basis states for the one-dimensional and two-dimensional irreducible representations, respectively. Similarly, for $\mathbf{Q} = \mathrm{K}$, we have $N_{\mathcal{G}_{\mathbf{Q}}} = 6$ and $d_{\xi_n} = 1$ for $\mathrm{K}_1$--$\mathrm{K}_6$, leading to 768 symmetry-adapted basis states. These numbers are in good agreement with the block sizes shown in Fig.~\ref{fig:2}, with a minor deviation attributed to the finite size of the basis. The exciton states obtained from diagonalization within each irreducible representation block coincide with those obtained through direct symmetry classification, confirming consistency between the two methods. The agreement of exciton irreducible representations with both the physical interpretation of compatibility relations and the block-diagonalization procedure establishes the robustness of our formalism. 
\begin{table*}
\caption[Excitonic irreducible representations at symmetry points and lines]{Irreducible representations of the first eight excitonic states at high-symmetry points and along high-symmetry lines in the Brillouin zone of monolayer MoS$_2$.}
\label{tab:table1}
\setlength\tabcolsep{12pt} 
\begin{tabular*}{\linewidth}{@{\extracolsep{\fill}} c c c c @{}}
\toprule
\shortstack{Symmetry \\ line} & 
\shortstack{Symmetry \\ points} &
\shortstack{Symmetry \\ group} &
\shortstack{Irreducible representations of the excitonic states} \\
\midrule
&$\mathrm{\Gamma}$ & $D_{3h}$ & $\Gamma_3 \oplus\Gamma_4 \oplus \Gamma_6 \oplus\Gamma_5 \oplus \Gamma_6$  \\
\midrule
$\mathrm{\Sigma} $&$\mathrm{\Sigma}_{pt_1}$ & $C_{2v}$ & $\mathrm{\Sigma}_3 \oplus \mathrm{\Sigma}_4 \oplus \mathrm{\Sigma}_2 \oplus \mathrm{\Sigma}_1 \oplus \mathrm{\Sigma}_3 \oplus \mathrm{\Sigma}_4 \oplus \mathrm{\Sigma}_2 \oplus \mathrm{\Sigma}_1$\\
&$\mathrm{\Sigma}_{pt_2}$ & $C_{2v}$ & $\mathrm{\Sigma}_3 \oplus \mathrm{\Sigma}_4 \oplus \mathrm{\Sigma}_2 \oplus \mathrm{\Sigma}_1 \oplus \mathrm{\Sigma}_2 \oplus \mathrm{\Sigma}_3 \oplus \mathrm{\Sigma}_4 \oplus \mathrm{\Sigma}_1$\\
\midrule
&$\mathrm{M}$ & $C_{2v}$ & $\mathrm{M}_3 \oplus \mathrm{M}_4 \oplus \mathrm{M}_2 \oplus \mathrm{M}_1 \oplus \mathrm{M}_2 \oplus \mathrm{M}_3 \oplus \mathrm{M}_4 \oplus \mathrm{M}_1$\\
\midrule
$\mathrm{T}$ &$\mathrm{T}_{pt_1}$ & $C_{s}$ & $\mathrm{T}_2 \oplus \mathrm{T}_2 \oplus \mathrm{T}_1 \oplus \mathrm{T}_1 \oplus \mathrm{T}_1 \oplus \mathrm{T}_2 \oplus \mathrm{T}_2 \oplus \mathrm{T}_2$\\
&$\mathrm{T}_{pt_2}$ & $C_{s}$ & $\mathrm{T}_2 \oplus \mathrm{T}_1 \oplus \mathrm{T}_2 \oplus \mathrm{T}_1 \oplus \mathrm{T}_2 \oplus \mathrm{T}_2 \oplus \mathrm{T}_1 \oplus \mathrm{T}_1$\\
\midrule
&$\mathrm{K}$ & $C_{3h}$ & $\mathrm{K}_6 \oplus \mathrm{K}_2 \oplus \mathrm{K}_4 \oplus \mathrm{K}_3 \oplus \mathrm{K}_5 \oplus \mathrm{K}_4 \oplus \mathrm{K}_2 \oplus \mathrm{K}_6$\\
\midrule
$\mathrm{\Lambda}$ &$\mathrm{\Lambda}_{pt_1}$ & $C_{s}$ & $\mathrm{\Lambda}_2 \oplus \mathrm{\Lambda}_1 \oplus \mathrm{\Lambda}_2 \oplus \mathrm{\Lambda}_1 \oplus \mathrm{\Lambda}_2 \oplus \mathrm{\Lambda}_2 \oplus \mathrm{\Lambda}_1 \oplus \mathrm{\Lambda}_1$\\
&$\mathrm{\Lambda}_{pt_2}$ & $C_{s}$ & $\mathrm{\Lambda}_1 \oplus \mathrm{\Lambda}_2 \oplus \mathrm{\Lambda}_2 \oplus \mathrm{\Lambda}_2 \oplus \mathrm{\Lambda}_1 \oplus \mathrm{\Lambda}_2 \oplus \mathrm{\Lambda}_1 \oplus \mathrm{\Lambda}_1$\\
&$\mathrm{\Lambda}_{pt_3}$ & $C_{s}$ & $\mathrm{\Lambda}_1 \oplus \mathrm{\Lambda}_2 \oplus \mathrm{\Lambda}_2 \oplus \mathrm{\Lambda}_1 \oplus \mathrm{\Lambda}_2 \oplus \mathrm{\Lambda}_2 \oplus \mathrm{\Lambda}_1 \oplus \mathrm{\Lambda}_1$\\
&$\mathrm{\Lambda}_{pt_4}$ & $C_{s}$ & $\mathrm{\Lambda}_2 \oplus \mathrm{\Lambda}_2 \oplus \mathrm{\Lambda}_1 \oplus \mathrm{\Lambda}_1 \oplus \mathrm{\Lambda}_2 \oplus \mathrm{\Lambda}_2 \oplus \mathrm{\Lambda}_1 \oplus \mathrm{\Lambda}_1$\\
\midrule
&$\mathrm{\Gamma}$ & $D_{3h}$ & $\Gamma_3 \oplus\Gamma_4 \oplus \Gamma_6 \oplus\Gamma_5 \oplus \Gamma_6$  \\
\bottomrule 
\end{tabular*}
\end{table*}

Symmetries of excitons can also be used to examine optical selection rules. The optical selection rule for a transition from an excitonic state $S$ at momentum $\mathbf{Q}$ to another excitonic state $S'$ at momentum $\mathbf{Q'}$ via a phonon mode $\nu$ at momentum $\mathbf{q}$ , involves the irreducible representations of excitons $\xi_{S,\mathbf{Q}}$, $\xi_{S',\mathbf{Q'}}$, and of the phonon $\xi_{\nu,\mathbf{q}}$, respectively \cite{PhysRevB.77.195201}. It is given as
\begin{equation*}
\xi_{S,\mathbf{Q}} \otimes \xi_{\nu,\mathbf{q}} \supset \xi_{S',\mathbf{Q'}}.
\end{equation*}
As an example, we show the selection rules for the optical transitions between exciton states of MoS$_2$ at $\mathbf{Q}=\mathbf{Q'}=0$ via a $\Gamma$-point phonon ($\mathbf{q}=0$). The selection rule becomes $\xi_{S,\mathbf{0}} \otimes \xi_{\nu,\mathbf{0}} \supset \xi_{S',\mathbf{0}}$. For both $\mathbf{Q}=0$ and $\mathbf{q}=0$, the little group is $D_{3h}$, identical to the crystal symmetry group. This approach is similar to the approach in Ref.~\cite{Ramanpaper}, where selection rules were formulated to analyze resonant Raman scattering in WSe$_{2}$/hBN heterostructures possessing $C_3$ symmetry at $\mathbf{q}=\mathbf{Q}=0$. For the MoS$_2$ case, the Kronecker product table between the irreducible representations corresponding to the initial excitonic state and the phonon modes is shown in Table~\ref{tab:table2}. However, not all irreducible representations appear at $\mathbf{q}=0$. The number of phonon modes for monolayer MoS$_2$ is 9  and these modes can be written as a direct sum of the irreducible representations as \cite{PhysRevB.84.155413,PhysRevB.91.235409}
\begin{equation*}
\Gamma_1  \oplus 2\Gamma_4 \oplus 2\Gamma_{6} \oplus \Gamma_5,
\end{equation*}
\captionsetup[table]{labelfont={color=black}}
\begin{table}[h!]
\centering
\caption{Kronecker product table between irreducible representations of $D_{3h}$}
\setlength{\tabcolsep}{4pt} 
\label{tab:table2}
\renewcommand{\arraystretch}{1.4} 
\begin{tabular}{c|cccccc}
$\xi_{S,\mathbf{0}} \otimes \xi_{\nu,\mathbf{0}}$ & $\Gamma_1$ & $\Gamma_2$ & $\Gamma_3$ & $\Gamma_4$ & $\Gamma_5$ & $\Gamma_6$ \\ \hline
$\Gamma_1$ & $\Gamma_1$ & $\Gamma_2$ & $\Gamma_3$ & $\Gamma_4$ & $\Gamma_5$ & $\Gamma_6$ \\
$\Gamma_2$ & $\Gamma_2$ & $\Gamma_1$ & $\Gamma_4$ & $\Gamma_3$ & $\Gamma_5$ & $\Gamma_6$ \\
$\Gamma_3$ & $\Gamma_3$ & $\Gamma_4$ & $\Gamma_1$ & $\Gamma_2$ & $\Gamma_6$ & $\Gamma_5$ \\
$\Gamma_4$ & $\Gamma_4$ & $\Gamma_3$ & $\Gamma_2$ & $\Gamma_1$ & $\Gamma_6$ & $\Gamma_5$ \\
$\Gamma_5$ & $\Gamma_5$ & $\Gamma_5$ & $\Gamma_6$ & $\Gamma_6$ & $\Gamma_1 \oplus \Gamma_2 \oplus \Gamma_6$ & $\Gamma_3 \oplus \Gamma_4 \oplus \Gamma_5$ \\
$\Gamma_6$ & $\Gamma_6$ & $\Gamma_6$ & $\Gamma_5$ & $\Gamma_5$ & $\Gamma_3 \oplus \Gamma_4 \oplus \Gamma_5$ & $\Gamma_1 \oplus \Gamma_2 \oplus \Gamma_6$ \\
\end{tabular}
\end{table}
The phonon modes in the direct sum of  irreducible representations are representated by a different notation in Ref.~\cite{PhysRevB.91.235409} as $A_{1}',A_{2}'',E'$, and $E''$, respectively. To define selection rules, as an example, we now consider the irreducible representations of the initial and final exciton states to be $\Gamma_5$ and $\Gamma_6$, respectively. These states correspond to the transitions from the doubly degenerate lowest B excition state to one of the doubly degenerate states in the A exciton. The following relations satisfy the selection rules:
\begin{eqnarray*}
\Gamma_5 \otimes \Gamma_3 &=& \Gamma_6\\
\Gamma_5 \otimes \Gamma_4 &=& \Gamma_6\\
\Gamma_5 \otimes \Gamma_5 &=& \Gamma_1 \oplus \Gamma_2 \oplus \Gamma_6 \supset \Gamma_6.
\end{eqnarray*}
As $\Gamma_3$ does not appear in the direct sum of phonon irreducible representations, only the transitions via phonons with  $\Gamma_4$ and $\Gamma_5$ are symmetry allowed, while transitions via $\Gamma_1$ and $\Gamma_6$ phonons are not symmetry allowed. Similarly, if one takes the initial and final states to be $\Gamma_5$ and $\Gamma_3$ or $\Gamma_4$, respectively, corresponding to the transitions from the doubly degenerate lowest B excition state to the other doubly degenerate state of the A exciton, the only symmetry allowed transition is via the $\Gamma_6$ phonon mode. This can be seen from the table as $\Gamma_5 \otimes \Gamma_6 = \Gamma_3 \oplus \Gamma_4 \oplus \Gamma_5 \supset \Gamma_3 \text{ and } \Gamma_4.$ This analysis can be further extended to study the symmetry based selection rules for the initial and final excitonic states at arbitrary center-of-mass momenta.

\section{\label{sec:level5}CONCLUSIONS}

In summary, in this paper we have established a general symmetry-based framework for excitons, incorporating both time-reversal and space-group operations. We showed how one can generate the excitonic states within the irreducible Brillouin zone and use space group symmetry operations to obtain the states at other points in the full Brillouin zone. This method allows great reduction in computational cost especially in problems where a fine sampling of the excitonic center-of-mass moementum is needed. Furthermore, by explicitly calculating the irreducible representations of the little groups and classifying excitonic states accordingly, we demonstrated how symmetry governs their degeneracies and band connectivities. Moreover, using projection operators, we constructed symmetry-adapted linear combinations of electron–hole product states, which block diagonalize the BSE Hamiltonian and provide a transparent symmetry classification of excitonic states. The irreducible representations of the excitonic states obtained with both the procedures were found to be in  agreement with those derived from compatibility relations, confirming the consistency of the formalism. This unified approach highlights the central role of symmetry in excitonic theory and provides a robust framework for analyzing optical selection rules and exciton band connectivities in a broad class of quantum materials.

\section{DATA AVAILABILITY}
The authors declare that the data supporting the findings of this study are availble within the main text and at \cite{dataset}. Other relevant data are available from the corresponding author upon request.

\section{\label{sec:level6}ACKNOWLEDGEMENTS}

We thank the Supercomputer Education and Research Centre (SERC) at the Indian Institute of Science (IISc) for providing the computational
facilities. R.B. acknowledges the funding from the Prime Minister’s Research Fellowship (PMRF), MHRD. M.J. and H.R.K. gratefully acknowledge the Nano mission of the Department of Science and Technology, India, and the Indian National Science Academy, India, for financial support under Grants No. DST/NM/TUE/QM-10/2019 and No. INSA/SP/SS/2023/, respectively.

\section*{APPENDIX A: Symmetry transformation of Hamiltonian}
We consider $\ket{n,\mathbf{k}}$ as the $n^{\mathrm{th}}$ electronic eigenstate at crystal momentum $\mathbf{k}$. Since $\hat{P}_{\{\mathcal{R}_{\mathbf{t}}|\mathbf{t}\}}$ commutes with $\hat{\mathcal{H}}$, we have
\begin{equation*}
    \hat{P}_{\{\mathcal{R}_{\mathbf{t}}|\mathbf{t}\}} \hat{\mathcal{H}} \ket{n,\mathbf{k}} 
    = \hat{\mathcal{H}} \hat{P}_{\{\mathcal{R}_{\mathbf{t}}|\mathbf{t}\}} \ket{n,\mathbf{k}}.
\end{equation*}
The left-hand side becomes
\begin{equation*}
    \hat{P}_{\{\mathcal{R}_{\mathbf{t}}|\mathbf{t}\}} \hat{\mathcal{H}} \ket{n,\mathbf{k}}
    = \hat{P}_{\{\mathcal{R}_{\mathbf{t}}|\mathbf{t}\}} \hat{\mathcal{H}}_{\mathbf{k}} \ket{n,\mathbf{k}},
\end{equation*}
while the right-hand side gives
\begin{align*}
    \hat{\mathcal{H}} \hat{P}_{\{\mathcal{R}_{\mathbf{t}}|\mathbf{t}\}} \ket{n,\mathbf{k}}
    = \hat{\mathcal{H}} \ket{n,\mathcal{R}_{\mathbf{t}}\mathbf{k}}
    &= \hat{\mathcal{H}}_{\mathcal{R}_{\mathbf{t}}\mathbf{k}}\ket{n,\mathcal{R}_{\mathbf{t}}\mathbf{k}}\\
    &= \hat{\mathcal{H}}_{\mathcal{R}_{\mathbf{t}}\mathbf{k}} 
      \hat{P}_{\{\mathcal{R}_{\mathbf{t}}|\mathbf{t}\}} \ket{n,\mathbf{k}}.
\end{align*}
Therefore,
\begin{equation*}
    \hat{P}_{\{\mathcal{R}_{\mathbf{t}}|\mathbf{t}\}} 
    \hat{\mathcal{H}}_{\mathbf{k}} \ket{n,\mathbf{k}}
    = \hat{\mathcal{H}}_{\mathcal{R}_{\mathbf{t}}\mathbf{k}} 
      \hat{P}_{\{\mathcal{R}_{\mathbf{t}}|\mathbf{t}\}} \ket{n,\mathbf{k}},
\end{equation*}
which leads to
\begin{equation*}
    \hat{P}_{\{\mathcal{R}_{\mathbf{t}}|\mathbf{t}\}} 
    \hat{\mathcal{H}}_{\mathbf{k}} 
    \left(\hat{P}_{\{\mathcal{R}_{\mathbf{t}}|\mathbf{t}\}}\right)^{-1}
    = \hat{\mathcal{H}}_{\mathcal{R}_{\mathbf{t}}\mathbf{k}}.
\end{equation*}

\section*{APPENDIX B: Translational symmetry in Bethe-Salpeter Equation (BSE) Hamiltonian}

\subsection*{A1. BSE Hamiltonian in real space}
We begin with the Bethe-Salpeter equation (BSE) for the electron-hole amplitude 
\(\Psi(\mathbf{r}_e, \mathbf{r}_h)\), where \(\mathbf{r}_e\) and \(\mathbf{r}_h\) are the electron and hole coordinates, respectively.
In real space, the BSE Hamiltonian acts as a four-point kernel:
\begin{equation}
\label{eq:BSE-real}
\begin{aligned}
&\mathcal{H}_{\mathrm{BSE}}(\mathbf{r}_e,\mathbf{r}_h;\,\mathbf{r}_e',\mathbf{r}_h') = 
\mathcal{H}_e(\mathbf{r}_e, \mathbf{r}_e')\,\delta(\mathbf{r}_h , \mathbf{r}_h') 
\\& + \mathcal{H}_h(\mathbf{r}_h, \mathbf{r}_h')\,\delta(\mathbf{r}_e - \mathbf{r}_e')
 -\,W(\mathbf{r}_e, \mathbf{r}_h)\,\delta(\mathbf{r}_e - \mathbf{r}_e')\,\delta(\mathbf{r}_h - \mathbf{r}_h') \\
& +\,v(\mathbf{r}_e, \mathbf{r}_h')\,\delta(\mathbf{r}_e - \mathbf{r}_h)\,\delta(\mathbf{r}_e' - \mathbf{r}_h')
\end{aligned}
\end{equation}
where \(\mathcal{H}_e(\mathbf{r}_e, \mathbf{r}_e')\) and \(\mathcal{H}_h(\mathbf{r}_h, \mathbf{r}_h')\) are the electron quasiparticle and hole quasiparticle parts of the Hamiltonian.
\(W(\mathbf{r}_e, \mathbf{r}_h)\) is the statically screened direct electron-hole interaction and \(v(\mathbf{r}_e, \mathbf{r}_h')\) is the bare Coulomb exchange term.

\subsection*{A2. Translational Invariance}
In periodic crystals, the underlying crystal, ionic potentials, the Coulomb interactions, quasiparticle self-energy (in the GW approximation), and screening are invariant under translations by any Bravais lattice vector \(\mathbf{R}\). Also, the one-particle lattice translation operator, \(\hat{T}^{e/h}_{\mathbf{R}}\) commutes the one-particle Hamiltonians, \(\mathcal{H}_{e/h}(\mathbf{r},\mathbf{r}')\). These properties leads to the following
\begin{enumerate}
  \item The electron and hole quasiparticle Hamiltonians are invariant with respect to discrete lattice translations:
  \begin{equation}
  \mathcal{H}_{e/h}(\mathbf{r}+\mathbf{R},\mathbf{r}'+\mathbf{R})=\mathcal{H}_{e/h}(\mathbf{r},\mathbf{r}')
  \end{equation}
  \item The interactions are invariant with respect to the discrete lattice translations as listed below:
  \begin{align}
W(\mathbf{r}+\mathbf{R},\mathbf{r}'+\mathbf{R})&=W(\mathbf{r},\mathbf{r}')
\notag\\
v(\mathbf{r}+\mathbf{R},\mathbf{r}'+\mathbf{R})&=v(\mathbf{r},\mathbf{r}')
  \end{align}
\end{enumerate}
\subsection*{A3. BSE Kernel invariance under simultaneous translations}
\noindent
We translate all four coordinates by the same Bravais vector \(\mathbf{R}\) as follows:
\[
(\mathbf{r}_e,\mathbf{r}_h,\mathbf{r}'_e,\mathbf{r}'_h)\mapsto(\mathbf{r}_e+\mathbf{R},\mathbf{r}_h+\mathbf{R},\mathbf{r}'_e+\mathbf{R},\mathbf{r}'_h+\mathbf{R})
\]
We examine each term in \(H\):
\begin{equation}
\begin{aligned}
&\mathcal{H}_{e}(\mathbf{r}_e+\mathbf{R},\mathbf{r}'_e+\mathbf{R})\,\delta(\mathbf{r}_h+\mathbf{R}-\mathbf{r}'_h-\mathbf{R})
\\
&= \mathcal{H}_{e}(\mathbf{r}_e,\mathbf{r}'_e)\,\delta(\mathbf{r}_h-\mathbf{r}'_h)
\\[4pt]
&\mathcal{H}_{h}(\mathbf{r}_h+\mathbf{R},\mathbf{r}'_h+\mathbf{R})\,\delta(\mathbf{r}_e+\mathbf{R}-\mathbf{r}'_e-\mathbf{R})
\\
&= \mathcal{H}_{h}(\mathbf{r}_h,\mathbf{r}'_h)\,\delta(\mathbf{r}_e-\mathbf{r}'_e),
\\[4pt]
&W(\mathbf{r}_e+\mathbf{R},\mathbf{r}_h+\mathbf{R})\,
\delta(\mathbf{r}_e+\mathbf{R}-\mathbf{r}'_e-\mathbf{R})\,
\\
&\times\delta(\mathbf{r}_h+\mathbf{R}-\mathbf{r}'_h-\mathbf{R})
\\
&= W(\mathbf{r}_e,\mathbf{r}_h)\,\delta(\mathbf{r}_e-\mathbf{r}'_e)\,\delta(\mathbf{r}_h-\mathbf{r}'_h),
\\[4pt]
&v(\mathbf{r}_e+\mathbf{R},\mathbf{r}'_h+\mathbf{R})\,
\delta(\mathbf{r}_e+\mathbf{R}-\mathbf{r}_h+\mathbf{R})\,
\\
&\times\delta(\mathbf{r}'_e+\mathbf{R}-\mathbf{r}'_h+\mathbf{R})
\\
&= v(\mathbf{r}_e,\mathbf{r}'_h)\,\delta(\mathbf{r}_e-\mathbf{r}_h)\,\delta(\mathbf{r}'_e-\mathbf{r}'_h).
\end{aligned}
\end{equation}
Each term reproduces its unshifted form; therefore, 
\begin{align}
 &\mathcal{H}_{\mathrm{BSE}}(\mathbf{r}_e+\mathbf{R},\mathbf{r}_h+\mathbf{R};\,\mathbf{r}'_e+\mathbf{R},\mathbf{r}'_h+\mathbf{R}) \notag \\
&= \mathcal{H}_{\mathrm{BSE}}(\mathbf{r}_e,\mathbf{r}_h;\,\mathbf{r}'_e,\mathbf{r}'_h) 
\end{align}
This shows that the BSE Hamiltonian is invariant under simultaneous translations of all coordinates by any Bravais vector \(\mathbf{R}\).

\subsection*{A4. Translation operator commutes with the Hamiltonian}

We define the two-particle translation operator \(\hat{T}^{ex}_{\mathbf{R}}\) acting on a two-particle function, \(\Psi(\mathbf{r}_1,\mathbf{r}_2)\), as
\begin{equation}
\label{eq:translation}
\big[\hat{T}^{ex}_{\mathbf{R}} \Psi\big](\mathbf{r}_1,\mathbf{r}_2) 
= \Psi(\mathbf{r}_1 - \mathbf{R}, \mathbf{r}_2 - \mathbf{R})
\end{equation}
Applying the BSE Hamiltonian to the translated amplitude gives
\begin{equation}
\begin{aligned}
&\left[ \hat{\mathcal{H}}_{\mathrm{BSE}}\, \hat{T}^{ex}_{\mathbf{R}} \Psi \right](\mathbf{r}_1,\mathbf{r}_2) \\
&= \iint d\mathbf{r}_1' d\mathbf{r}_2' \;
\mathcal{H}_{\mathrm{BSE}}(\mathbf{r}_1,\mathbf{r}_2;\,\mathbf{r}_1',\mathbf{r}_2') \,
\Psi(\mathbf{r}_1' - \mathbf{R}, \mathbf{r}_2' - \mathbf{R}).
\\
&= \iint d\mathbf{r}_1' d\mathbf{r}_2' \;
\mathcal{H}_{\mathrm{BSE}}(\mathbf{r}_1 - \mathbf{R},\mathbf{r}_2 - \mathbf{R};\,\mathbf{r}_1' - \mathbf{R},\mathbf{r}_2' - \mathbf{R}) \,
\\
& \times\Psi(\mathbf{r}_1' - \mathbf{R}, \mathbf{r}_2' - \mathbf{R})
\end{aligned}
\end{equation}
In the last equation, we have used the translational invariance of the Hamiltonian. Changing integration variables to \(\mathbf{r}_1'' = \mathbf{r}_1' - \mathbf{R}\), \(\mathbf{r}_2'' = \mathbf{r}_2' - \mathbf{R}\), one finds:
\begin{equation}
\begin{aligned}
&\left[ \hat{\mathcal{H}}_{\mathrm{BSE}}\, \hat{T}^{ex}_{\mathbf{R}} \Psi \right](\mathbf{r}_1,\mathbf{r}_2) \\
&= \iint d\mathbf{r}_1'' d\mathbf{r}_2'' \;
\mathcal{H}_{\mathrm{BSE}}(\mathbf{r}_1 - \mathbf{R},\mathbf{r}_2 - \mathbf{R};\,\mathbf{r}_1'' ,\mathbf{r}_2'') \,
\Psi(\mathbf{r}_1'', \mathbf{r}_2'')
\\
&= \iint d\mathbf{r}_1'' d\mathbf{r}_2'' \;
\left[\hat{T}_{\mathbf{R}}\mathcal{H}_{\mathrm{BSE}}(\mathbf{r}_1,\mathbf{r}_2;\,\mathbf{r}_1'' ,\mathbf{r}_2'')\right] \,
\Psi(\mathbf{r}_1'', \mathbf{r}_2'')
\\
&= \left[\hat{T}^{ex}_{\mathbf{R}}\hat{\mathcal{H}}_{\mathrm{BSE}} \Psi \right](\mathbf{r}_1,\mathbf{r}_2)
\end{aligned}
\end{equation}
This leads to 
\begin{equation}
\hat{\mathcal{H}}_{\mathrm{BSE}}\,\hat{T}^{ex}_{\mathbf{R}} = \hat{T}^{ex}_{\mathbf{R}}\,\hat{\mathcal{H}}_{\mathrm{BSE}}.
\end{equation}
Therefore, \(\hat{\mathcal{H}}_{\mathrm{BSE}}\) commutes with the two-particle translation operator.
\subsection*{A5. Bloch's Theorem for the Electron-Hole Amplitude}
Since \(\hat{\mathcal{H}}_{\mathrm{BSE}}\) commutes with \(\hat{T}^{ex}_{\mathbf{R}}\), the electron-hole amplitudes are  eigenfunctions of both \(\hat{T}^{ex}_{\mathbf{R}}\) and \(\mathcal{H}_{\mathrm{BSE}}\).  Specifically,
\begin{equation}\label{eq:translation_property}
\hat{T}^{ex}_{\mathbf{R}} \Psi_{S,\mathbf{Q}}(\mathbf{r}_e,\mathbf{r}_h) 
= e^{i\mathbf{Q}\cdot\mathbf{R}} \Psi_{S,\mathbf{Q}}(\mathbf{r}_e,\mathbf{r}_h)
\end{equation}
where \(\mathbf{Q}\) is the total momentum of the two-particle excitation state and $S$ is the state index. Thus translational invariance ensures that two-particle excitaions can be labeled by a well-defined crystal momentum \(\mathbf{Q}\).

Now, we demonstrate the Bloch periodicity in the product state basis used in the main text. We begin from the electron-hole amplitude expansion of \(\Psi_{S,\mathbf Q}(\mathbf r_e,\mathbf r_h)\) [Eq. \ref{eq:excwavefn}] by simultaneously translating the electron and hole coordinates by $\mathbf R$, {\em{i.e.}}, \(
\mathbf r_e\mapsto\mathbf r_e+\mathbf R\) and \(\mathbf r_h\mapsto\mathbf r_h+\mathbf R\) which leaves the relative coordinate $\mathbf r$ unchanged and shifts the center-of-mass coordinate as \(
\mathbf R_{\mathrm{cm}}\mapsto \mathbf R_{\mathrm{cm}}+\mathbf R\). Using the single-particle Bloch theorem and evaluating amplitude at the translated arguments gives:
\begin{align}
&\Psi_{S,\mathbf Q}(\mathbf r_e+\mathbf R,\mathbf r_h+\mathbf R)
\notag \\
&=\sum_{v,c,\mathbf k} A^{S,\mathbf Q}_{v,c,\mathbf k}\;
\Phi_{c,\mathbf k}(\mathbf r_e+\mathbf R)\,\Phi_{v,\mathbf k-\mathbf Q}^*(\mathbf r_h+\mathbf R)
\notag \\
&=\sum_{v,c,\mathbf k} A^{S,\mathbf Q}_{v,c,\mathbf k}\;
e^{i\mathbf k\cdot\mathbf R}\Phi_{c,\mathbf k}(\mathbf r_e)\;
\bigl(e^{i(\mathbf k-\mathbf Q)\cdot\mathbf R}\Phi_{v,\mathbf k-\mathbf Q}(\mathbf r_h)\bigr)^*
\notag \\
&=e^{i\mathbf Q\cdot\mathbf R}\sum_{v,c,\mathbf k} A^{S,\mathbf Q}_{v,c,\mathbf k}\;
\Phi_{c,\mathbf k}(\mathbf r_e)\,\Phi_{v,\mathbf k-\mathbf Q}^*(\mathbf r_h)
\notag \\
&= e^{i\mathbf Q\cdot\mathbf R}\,\Psi_{S,\mathbf Q}(\mathbf r_e,\mathbf r_h)
\end{align}
In center-of-mass and relative coordinates this reads
\begin{equation}
\Psi_{S,\mathbf Q}(\mathbf R_{\mathrm{cm}}+\mathbf R,\mathbf r)
= e^{\,i\mathbf Q\cdot\mathbf R}\;\Psi_{S,\mathbf Q}(\mathbf R_{\mathrm{cm}},\mathbf r)
\end{equation}
We define
\[
\mathbf R_{\mathrm{cm}} = \alpha\,\mathbf r_e + \beta\,\mathbf r_h, 
\quad \mathbf r = \mathbf r_e - \mathbf r_h, 
\quad \alpha + \beta = 1
\]
so that 
\(\mathbf r_e = \mathbf R_{\mathrm{cm}} + \beta\mathbf r\),
\(\mathbf r_h = \mathbf R_{\mathrm{cm}} - \alpha\mathbf r\), \(\mathbf k_h = \mathbf k - \beta\mathbf Q\) and \(\mathbf k_e = \mathbf k + \alpha\mathbf Q\) 

\begin{align}
&\Psi_{S,\mathbf Q}(\mathbf R_{\mathrm{cm}},\mathbf r)
= e^{i\mathbf Q\cdot\mathbf R_{\mathrm{cm}}}
\sum_{v,c,\mathbf k} A^{S,\mathbf Q}_{v,c,\mathbf k+\alpha\mathbf Q} \;
e^{i\mathbf k\cdot\mathbf r} 
\notag \\
& \times u_{c,\mathbf k+\alpha\mathbf Q}(\mathbf R_{\mathrm{cm}}+\beta\mathbf r) \;
u^{*}_{v,\mathbf k-\beta\mathbf Q}(\mathbf R_{\mathrm{cm}}-\alpha\mathbf r)
\end{align}
Including normalization factors, the Bloch periodic form of the electron-hole amplitude can then be written as
\begin{equation}
\Psi_{S,\mathbf Q}(\mathbf R_{\mathrm{cm}},\mathbf r)
= \frac{1}{\sqrt{N_Q}}\,e^{i\mathbf Q\cdot\mathbf R_{\mathrm{cm}}}
F_{S,\mathbf Q}(\mathbf R_{\mathrm{cm}},\mathbf r)
\end{equation}
with the cell-periodic part
\begin{align}
&F_{S,\mathbf Q}(\mathbf R_{\mathrm{cm}},\mathbf r)
= \frac{1}{\sqrt{N_k}}
\sum_{v,c,\mathbf k} A^{S,\mathbf Q}_{v,c,\mathbf k+\alpha\mathbf Q} \;
e^{i\mathbf k\cdot\mathbf r} \;
\notag \\
& \times u_{c,\mathbf k+\alpha\mathbf Q}(\mathbf R_{\mathrm{cm}}+\beta\mathbf r) \;
u^{*}_{v,\mathbf k-\beta\mathbf Q}(\mathbf R_{\mathrm{cm}}-\alpha\mathbf r)
\end{align}

Here, the cell-periodic part of two-particle excitation is lattice periodic in the center-of-mass coordinate:
\begin{equation}
F_{S,\mathbf Q}(\mathbf R_{\mathrm{cm}}+\mathbf R,\mathbf r)=F_{S,\mathbf Q}(\mathbf R_{\mathrm{cm}},\mathbf r).
\end{equation}

For the symmetric case \(\alpha=\beta=\frac12\), the conduction and valence states carry
momenta \(\mathbf k+\frac12\mathbf Q\) and \(\mathbf k-\frac12\mathbf Q\), respectively. For the calculation within \textsc{BerkeleyGW}, the form used in Eq. \ref{eq:excwavefn} is used, which is \(\alpha=0\) and \(\beta=1\). The Bloch-periodic form expressed in these coordinates was previously discussed in Ref. \cite{PhysRevB.108.125118} and is included here for completeness.

\section*{APPENDIX C: Transformation of excitons under time-reversal symmetry}

Let \(\Theta\) denote the time-reversal operator acting on the exciton amplitude
\(\Psi_{S,\mathbf{Q}}(\mathbf r_e,\mathbf r_h)\). Its action is defined as
\begin{equation}
\left(\hat{P}^{ex}_\Theta \Psi_{S,\mathbf Q}\right)(\mathbf r_e, \mathbf r_h)
= \Psi^*_{S,\mathbf Q}(\mathbf r_e, \mathbf r_h)
\end{equation}
where the complex conjugation acts on the coefficients and the single-particle spinor parts
of the electron and hole wave functions. Since time-reversal reverses all crystal momenta,
we have, at the single-particle operator level,
\begin{equation}
\hat{P}^{ex}_\Theta\, b^\dagger_{n,\mathbf{k},s}\, \hat{P}^{ex\,-1}_\Theta
= \sum_{s'} \left(i\sigma_y\right)_{s s'}\, b^\dagger_{n,-\mathbf{k},s'}
\end{equation}
and similarly for the hole creation operators; the same transformation is inherited by
the exciton amplitude coefficients in the Bloch basis.

Using the translation property defined in Eqs. \ref{eq:translation} and
acting with \(\hat{P}^{ex}_\Theta\) on the translated wave function gives:
\begin{align}
&\left[\hat{P}^{ex}_\Theta\, \hat{T}^{ex}_{\mathbf{R}} \Psi_{S,\mathbf Q}\right](\mathbf r_e,\mathbf r_h) \notag \\
&= \hat{P}^{ex}_\Theta \bigg( \Psi_{S,\mathbf Q}(\mathbf r_e - \mathbf{R},\, \mathbf r_h - \mathbf{R}) \bigg) \notag \\
&= \Psi^*_{S,\mathbf Q}(\mathbf r_e - \mathbf{R},\, \mathbf r_h - \mathbf{R})
\end{align}
Because \(\hat{P}^{ex}_\Theta\) is antiunitary it complex-conjugates the phase factor,
so applying it to the translation property defined in Eq. \ref{eq:translation_property} yields
\begin{align}
\left[\hat{P}^{ex}_\Theta\, \hat{T}^{ex}_{\mathbf{R}} \Psi_{S,\mathbf Q}\right](\mathbf r_e,\mathbf r_h)
&= \left[\hat{P}^{ex}_\Theta \left( e^{i\mathbf{Q}\cdot\mathbf{R}} \Psi_{S,\mathbf Q} \right)\right](\mathbf r_e,\mathbf r_h) \notag \\
&= e^{-i\mathbf{Q}\cdot\mathbf{R}} \left[\hat{P}^{ex}_\Theta \Psi_{S,\mathbf Q}\right](\mathbf r_e,\mathbf r_h)
\end{align}
We compare this with the defining translation property similar to Eq. \ref{eq:translation_property} for the amplitudes at \(-\mathbf Q\):
\begin{equation}
\hat{T}^{ex}_{\mathbf{R}} \Psi_{S,-\mathbf Q}(\mathbf r_e,\mathbf r_h)
= e^{-i\mathbf{Q}\cdot\mathbf{R}} \Psi_{S,-\mathbf Q}(\mathbf r_e,\mathbf r_h)
\end{equation}
We therefore conclude that the time reversed exciton amplitude
\(\left[\hat{P}^{ex}_\Theta \Psi_{S,\mathbf Q}\right](\mathbf r_e,\mathbf r_h)\)
belongs to the momentum block \(-\mathbf Q\)
and using the fact that the time-reversal commutes with the BSE Hamiltonian, we get
\begin{equation}
\hat{P}^{ex}_\Theta\, \mathcal{H}_{\mathbf Q}\, \hat{P}^{ex\,-1}_\Theta
= \mathcal{H}_{-\mathbf Q}
\end{equation}
If \(\mathcal{H}_{\mathbf Q}\, \Psi_{S,\mathbf Q}(\mathbf r_e,\mathbf r_h)
= \Omega_{S,\mathbf Q}\, \Psi_{S,\mathbf Q}(\mathbf r_e,\mathbf r_h)\), then applying \(\hat{P}^{ex}_\Theta\) yields
\begin{align}
\mathcal{H}_{-\mathbf Q}\, \left[\hat{P}^{ex}_\Theta \Psi_{S,\mathbf Q}\right](\mathbf r_e,\mathbf r_h)
&= \left[\hat{P}^{ex}_\Theta \mathcal{H}_{\mathbf Q} \Psi_{S,\mathbf Q}\right](\mathbf r_e,\mathbf r_h) \notag\\
&= \Omega_{S,\mathbf Q}\, \left[\hat{P}^{ex}_\Theta \Psi_{S,\mathbf Q}\right](\mathbf r_e,\mathbf r_h)
\end{align}
Thus the time-reversed amplitude is an eigenfunction in the \(-\mathbf Q\) block with the same eigenvalue:
\begin{equation}
\Omega_{S,\mathbf Q} = \Omega_{S,-\mathbf Q}, \qquad
\left[\hat{P}^{ex}_\Theta \Psi_{S,\mathbf Q}\right](\mathbf r_e,\mathbf r_h)
= \Psi_{S,-\mathbf Q}(\mathbf r_e,\mathbf r_h)
\end{equation}
Equivalently, in Dirac notation,
\begin{equation}
\hat{P}^{ex}_\Theta \ket{S,\mathbf Q} = \ket{S,-\mathbf Q},
\end{equation}
which proves Eq.~\eqref{eq:timerevexc}.

\section*{APPENDIX D: Transformation of excitons under space group symmetries}

We consider a space group operation \(\{\mathcal{R}_{\mathbf{t}}|\mathbf{t}\}\). Its action on the exciton amplitude is defined by
\begin{equation}
\left(\hat{P}_{\{\mathcal{R}_{\mathbf{t}}|\mathbf{t}\}}^{ex} \Psi_{S,\mathbf{Q}}\right)(\mathbf r_e, \mathbf r_h)
= \Psi({\mathcal{R}_{\mathbf{t}}}^{-1}(\mathbf r_e - \mathbf{t}),\, {\mathcal{R}_{\mathbf{t}}}^{-1}(\mathbf r_h - \mathbf{t}))
\end{equation}
First, we show that \(\hat{P}_{\{\mathcal{R}_{\mathbf{t}}|\mathbf{t}\}}^{ex} \ket{S,\mathbf Q}\) belongs to the momentum block \(\mathcal{R}_{\mathbf{t}} \mathbf{Q}\).
We define the combined action of translations as defined in Eq. \ref{eq:translation} along with space group operations as
\begin{align}
&\hat{P}_{\{\mathcal{R}_{\mathbf{t}}|\mathbf{t}\}}^{ex}\, \hat{T}^{ex}_{\mathbf{R}}\, (\hat{P}_{\{\mathcal{R}_{\mathbf{t}}|\mathbf{t}\}}^{ex})^{-1} \Psi_{S,\mathbf{Q}}(\mathbf r_e, \mathbf r_h) 
\notag \\
&= \hat{T}^{ex}_{\mathbf{R}}\, (\hat{P}_{\{\mathcal{R}_{\mathbf{t}}|\mathbf{t}\}}^{ex})^{-1} \Psi_{S,\mathbf{Q}}({\mathcal{R}_{\mathbf{t}}}^{-1}(\mathbf r_e - \mathbf{t}),\, {\mathcal{R}_{\mathbf{t}}}^{-1}(\mathbf r_h - \mathbf{t})) 
\notag \\
&= \Psi_{S,\mathbf{Q}}\bigg(\mathcal{R}_{\mathbf{t}}({\mathcal{R}_{\mathbf{t}}}^{-1}(\mathbf r_e - \mathbf{t}) - \mathbf{R}) + \mathbf{t},
\notag \\
& \mathcal{R}_{\mathbf{t}}({\mathcal{R}_{\mathbf{t}}}^{-1}(\mathbf r_h - \mathbf{t}) - \mathbf{R}) + \mathbf{t}\bigg) 
\notag \\
&= \Psi_{S,\mathbf{Q}}(\mathbf r_e - \mathcal{R}_{\mathbf{t}} \mathbf{R},\, \mathbf r_h - \mathcal{R}_{\mathbf{t}} \mathbf{R}) \notag \\
&= \left(\hat{T}^{ex}_{\mathcal{R}_{\mathbf{t}} \mathbf{R}} \Psi_{S,\mathbf{Q}}\right)(\mathbf r_e, \mathbf r_h)
\end{align}

This gives the identity
\begin{equation}
\hat{P}_{\{\mathcal{R}_{\mathbf{t}}|\mathbf{t}\}}^{ex}\, \hat{T}^{ex}_{\mathbf{R}}\, (\hat{P}_{\{\mathcal{R}_{\mathbf{t}}|\mathbf{t}\}}^{ex})^{-1}
= \hat{T}^{ex}_{\mathcal{R}_{\mathbf{t}} \mathbf{R}}
\end{equation}
Now, using the translation property defined in Eq. \ref{eq:translation_property}, we evaluate
\begin{align}
&\left(\hat{T}^{ex}_{\mathbf{R}} \hat{P}_{\{\mathcal{R}_{\mathbf{t}}|\mathbf{t}\}}^{ex} \Psi_{S,\mathbf Q}\right)(\mathbf r_e, \mathbf r_h) 
\notag \\
&= \left(\hat{P}_{\{\mathcal{R}_{\mathbf{t}}|\mathbf{t}\}}^{ex} \hat{T}^{ex}_{{\mathcal{R}_{\mathbf{t}}}^{-1} \mathbf{R}} \Psi_{S,\mathbf Q} \right)(\mathbf r_e, \mathbf r_h) 
\notag \\
&= e^{i \mathbf Q \cdot {\mathcal{R}_{\mathbf{t}}}^{-1} \mathbf{R}} \left(\hat{P}_{\{\mathcal{R}_{\mathbf{t}}|\mathbf{t}\}}^{ex} \Psi_{S,\mathbf Q} \right)(\mathbf r_e, \mathbf r_h) 
\notag \\
&= e^{i (\mathcal{R}_{\mathbf{t}} \mathbf{Q}) \cdot \mathbf{R}} \left(\hat{P}_{\{\mathcal{R}_{\mathbf{t}}|\mathbf{t}\}}^{ex} \Psi_{S,\mathbf Q} \right)(\mathbf r_e, \mathbf r_h)
\end{align}
which implies that \(\hat{P}_{\{\mathcal{R}_{\mathbf{t}}|\mathbf{t}\}}^{ex} \ket{S,\mathbf Q}\) lies in the momentum block labeled by \(\mathcal{R}_{\mathbf{t}} \mathbf{Q}\). Also, the BSE Hamiltonian commutes with the symmetry operation. Therefore,
\begin{equation}
\mathcal H_{\mathcal{R}_{\mathbf{t}} \mathbf{Q}}\, \hat{P}_{\{\mathcal{R}_{\mathbf{t}}|\mathbf{t}\}}^{ex} = \hat{P}_{\{\mathcal{R}_{\mathbf{t}}|\mathbf{t}\}}^{ex}\, \mathcal H_\mathbf Q
\end{equation}
If \(\mathcal H_\mathbf Q\ket{S,\mathbf{Q}} = \Omega_{S,\mathbf Q}\ket{S,\mathbf{Q}}\), then
\begin{align}
&\left[\hat{P}_{\{\mathcal{R}_{\mathbf{t}}|\mathbf{t}\}}^{ex}\mathcal H_\mathbf Q\right]\ket{S,\mathbf{Q}} = \Omega_{S,\mathbf Q}\left[\hat{P}_{\{\mathcal{R}_{\mathbf{t}}|\mathbf{t}\}}^{ex}\ket{S,\mathbf{Q}}\right] \notag\\
&\mathcal H_{\mathcal{R}_{\mathbf{t}} \mathbf{Q}}\left[\hat{P}_{\{\mathcal{R}_{\mathbf{t}}|\mathbf{t}\}}^{ex}\ket{S,\mathbf{Q}}\right] = \Omega_{S,\mathbf Q}\left[\hat{P}_{\{\mathcal{R}_{\mathbf{t}}|\mathbf{t}\}}^{ex}\ket{S,\mathbf{Q}}\right]
\end{align}

Hence the rotated state \(\hat{P}_{\{\mathcal{R}_{\mathbf{t}}|\mathbf{t}\}}^{ex} \ket{S,\mathbf Q}\) is an eigenstate of \(\mathcal{R}_{\mathbf{t}} \mathbf{Q}\) momentum block of the Hamiltonian with the same eigenvalue. So, the following holds:
\begin{equation}
\Omega_{S,\mathcal{R}_{\mathbf{t}} \mathbf{Q}} = \Omega_{S,\mathbf Q}, \qquad
\ket{S,\mathcal{R}_{\mathbf{t}} \mathbf{Q}} = \hat{P}_{\{\mathcal{R}_{\mathbf{t}}|\mathbf{t}\}}^{ex} \ket{S,\mathbf Q}
\end{equation}


%

\end{document}